\documentclass[journal,10pt]{IEEEtran}
\IEEEoverridecommandlockouts
\usepackage{cite}
\usepackage{url} 
\usepackage{amsmath,amssymb,amsfonts}
\usepackage{algorithm}
\usepackage{algorithmic}
\usepackage{graphicx}
\usepackage{subcaption}
\usepackage{textcomp}
\def\BibTeX{{\rm B\kern-.05em{\sc i\kern-.025em b}\kern-.08em
		T\kern-.1667em\lower.7ex\hbox{E}\kern-.125emX}}    

\usepackage{amsmath,amssymb,amsfonts,mathrsfs,amsthm,graphicx,epstopdf,bm}
\usepackage{esint,epsfig,soul} 
\usepackage{cite, color,kotex, comment}
\usepackage[shortlabels]{enumitem}
\usepackage[usenames,dvipsnames,svgnames]{xcolor}
\usepackage{bbm}
\usepackage{multirow}

\newtheorem{theorem}{Theorem}

\newtheorem{lemma}{Lemma}
\newtheorem{proposition}{Proposition}

\theoremstyle{definition}

\newtheorem{assumption}{Assumption}

\newcommand{\gray}[1]{{\color{gray}{#1}}}
\renewcommand{\gray}[1]{}

\newcommand{\romanum}[1]{\uppercase\expandafter{\romannumeral#1\relax}}

\begin{document}
	
	\title{
    Communication-Efficient Approximate Gradient Coding for Distributed Learning in Heterogeneous Systems 
	}
	\author{Heekang~Song,~\IEEEmembership{Member,~IEEE},
		and~Wan~Choi,~\IEEEmembership{Fellow,~IEEE}
		\thanks{
			H. Song is with the Institute of New Media and Communications, SNU, Seoul 08826, Korea (e-mail: hghsong95@gmail.com). 
		}
		\thanks{	
			W.~Choi is with  the Department of Electrical and Computer Engineering, Seoul National University (SNU), and  the Institute of New Media and Communications, SNU, Seoul 08826, Korea (e-mail:   wanchoi@snu.ac.kr). (\emph{Corresponding author: Wan Choi}).
		}
	}
	\maketitle
	
	\begin{abstract}
    We propose a communication-efficient optimally structured gradient coding scheme to jointly address straggler resilience and communication efficiency in heterogeneous distributed learning. By establishing a unified framework that simultaneously optimizes gradient coding and quantization, we formulate an optimization problem to minimize residual error subject to an unbiasedness constraint. We rigorously establish the joint global optimum by deriving a closed-form code structure coupled with an optimal bit allocation strategy, while simultaneously proposing a low-complexity bit allocation algorithm that efficiently yields near-optimal performance. We provide rigorous convergence analysis for convex and smooth functions. Experiments on the COCO dataset demonstrate that our joint design significantly accelerates convergence and enhances communication efficiency compared to existing baselines.
	\end{abstract}
	
	\section{Introduction}

    \IEEEPARstart{I}{N} recent years, the rapid advancements in deep learning have underscored the significance of large datasets and large-scale AI models as critical components for performance enhancement. Breakthrough models have not only demonstrated unprecedented capabilities but have also transformed the industrial landscape, reshaping how AI technologies are applied across various domains. However, training these massive models demands immense computational resources that often exceed the capacity of a single facility. Consequently, the scope of model training has expanded beyond individual clusters to cross-datacenter environments, where resources from geographically distributed infrastructures are aggregated to overcome local computation bottlenecks.

    To support this scale, distributed computing has emerged as the fundamental solution. Building on the success of foundational frameworks like MapReduce \cite{ref:MapReduce} and Spark \cite{ref:Spark}, distributed computing has evolved to address the complexities of modern AI, solidifying its role in mitigating computation and communication constraints \cite{ref:CMC, ref:CDC}. This evolution is particularly evident in the rise of distributed learning, which now frequently operates in cross-datacenter settings to accommodate the growing need for efficient, global-scale training.
    
    A typical distributed learning architecture consists of a central coordinator (master node) that minimizes a loss function using gradient-based optimization (e.g., Gradient Descent). The master node partitions the dataset and distributes tasks to worker nodes, which compute local gradients in parallel. The master then aggregates these gradients to update the model parameters. While this workload distribution enhances scalability, the cross-datacenter environment inherently introduces significant heterogeneity in computational speeds and network latencies. In such settings, the overall system performance is frequently dictated by the slowest node—known as a \emph{straggler}—which creates severe bottlenecks and limits the efficiency of the training process.
    
    To mitigate the impact of stragglers in distributed learning, \emph{Gradient Coding (GC)} has been extensively studied as a prominent solution \cite{ref:GC, ref:GC2, ref:GC3, ref:AGC, ref:AGC2, ref:AGC3, ref:AGC4, ref:SGC}. Pioneered by \cite{ref:GC}, this technique exploits data replication to encode partial gradients, enabling the exact reconstruction of the full gradient even in the presence of stragglers. While subsequent work \cite{ref:GC2} explored the trade-off between communication overhead and straggler resilience, these exact recovery methods typically rely on the unrealistic assumption of a known, fixed number of stragglers and necessitate high data redundancy, imposing a significant computational burden on worker nodes.
    
    To address these limitations, \emph{Approximate Gradient Coding (AGC)} has emerged as a practical alternative, operating under more realistic probabilistic straggler models \cite{ref:GC3, ref:AGC, ref:AGC2, ref:AGC3, ref:AGC4, ref:SGC}. Unlike exact GC, AGC relaxes the requirement for perfect reconstruction, instead utilizing an estimated gradient sum for model updates. This relaxation is well-justified, as stochastic optimization algorithms like SGD are inherently robust to noise and approximation errors. By minimizing the residual error between the true and approximate gradients rather than enforcing exact recovery, AGC significantly reduces the computational overhead associated with data replication.
    
    Various strategies have been proposed to construct efficient AGC schemes. For instance, \cite{ref:GC3} utilized the normalized adjacency matrix of expander graphs, while \cite{ref:AGC} introduced a scheme based on sparse random graphs with Bernoulli sampling. The properties of Fractional Repetition (FR) codes were analyzed in \cite{ref:AGC2}, although their application is constrained by specific divisibility requirements between nodes and replication factors. On the theoretical front, \cite{ref:AGC3} characterized the fundamental trade-offs among replication factor, straggler count, and residual error. Building on these foundations, \cite{ref:AGC4} recently proposed a dynamic decoding framework using expander graphs to further minimize residual error during training.

    While previous studies \cite{ref:GC3, ref:AGC, ref:AGC2, ref:AGC3, ref:AGC4} primarily focused on minimizing the residual error between the estimated and true gradients, this metric alone does not necessarily guarantee model convergence. Addressing this gap, Stochastic Gradient Coding (SGC) \cite{ref:SGC} introduced a pairwise data distribution scheme to construct an unbiased gradient estimator, ensuring robust convergence even under severe straggler conditions where residual-error-focused methods often falter. However, SGC’s reliance on strict pairwise constraints can significantly increase implementation complexity depending on the system scale.

    Furthermore, given the iterative aggregation process in distributed learning, communication efficiency is as critical as computation. Yet, most existing works, including SGC, overlook the transmission overhead. A notable exception is 1-bit SGC \cite{ref:SGC2}, which integrated 1-bit quantization into the SGC framework. However, this approach merely applies quantization sequentially to the straggler mitigation scheme, failing to analyze the joint impact of stragglers and quantization noise. Moreover, being restricted to extreme 1-bit quantization, it inevitably slows down convergence and fails to address the general trade-off between communication volume and convergence speed.

    In summary, previous studies have generally pursued two directions: minimizing residual error or ensuring unbiasedness via specific coding designs. However, most prior research relies on unrealistic homogeneous straggler assumptions, which can lead to dataset neglect and poor generalization in practical, heterogeneous settings. To address these issues, our previous work on AGC in heterogeneous systems \cite{ref:HAGC} proposed a novel technique specifically designed to enhance stability and convergence in heterogeneous environments. Building on this foundation, this work extends the analysis to improve communication efficiency in heterogeneous systems by jointly incorporating gradient coding and quantization, a critical challenge not addressed in \cite{ref:HAGC}. To the best of our knowledge, this is the first unified framework to jointly optimizes gradient coding and quantization for heterogeneous systems. Unlike prior attempts that simply apply limited 1-bit quantization sequentially, our approach integrates straggler mitigation with a flexible quantization scheme. This joint design maximizes both straggler resilience and communication efficiency simultaneously, adapting to any bit budget.
    
	\section{Preliminaries} \label{sec:PRE}

\subsection{Distributed Learning with Gradient Coding}
We consider a distributed learning framework with one master node and $k$ worker nodes. 
Given a dataset $\mathcal{D}=\{\mathcal{D}_1,\ldots,\mathcal{D}_n\}$ partitioned into $n$ data blocks, the master aims to find $\boldsymbol{\beta}^*\in\mathbb{R}^l$ that minimizes $\sum_{j=1}^n L(\mathcal{D}_j,\boldsymbol{\beta})$. 
At iteration $t$, let $g_j^{(t)}=\nabla L(\mathcal{D}_j,\boldsymbol{\beta}_t)$ denote the gradient of partition $\mathcal{D}_j$, and let $g^{(t)}=\sum_{j=1}^n g_j^{(t)}$ be the full gradient. 
The ideal gradient descent update is
\begin{equation} \label{update_param}
    \boldsymbol{\beta}_{t+1} = \boldsymbol{\beta}_t - \gamma_t g^{(t)},
\end{equation}
where $\gamma_t$ is the learning rate. 
In distributed learning, however, the master estimates $g^{(t)}$ from the responses of worker nodes, some of which may become stragglers. 
Gradient coding mitigates this issue by assigning possibly overlapping data batches $\mathcal{B}_i\subseteq\mathcal{D}$ to workers and linearly encoding their computed partial gradients.

The procedure consists of data distribution, local computation, and gradient update. 
In the data distribution phase, each partition $\mathcal{D}_j$ is replicated across $d_j$ workers, and the average computation load is defined as $d=\frac{1}{n}\sum_{j=1}^n d_j$. 
This phase is performed once before training, while local computation and gradient update are repeated at every iteration. 
Given its assigned batch $\mathcal{B}_i$, worker $i$ computes the corresponding partial gradients and forms the encoded message $f_i^{(t)}=\sum_{\mathcal{D}_j\in\mathcal{B}_i}a_{i,j}g_j^{(t)}$, where $a_{i,j}\in\mathbb{R}$ is the encoding coefficient.
To reduce communication overhead, worker $i$ transmits the quantized message $q_i^{(t)}=Q_i(f_i^{(t)})$, where $Q_i(\cdot)$ is a stochastic quantizer. 
The master aggregates the received messages as
\begin{equation}
    \hat{g}^{(t)} = \sum_{i=1}^k \mathbb{I}_i w_i q_i^{(t)},
\end{equation}
where $w_i \in \mathbb{R}$ represents the decoding coefficient for worker node $i$. Here, $\mathbb{I}_i$ serves as an indicator variable that takes the value $1$ if worker node $i$ is a non-straggler and $0$ otherwise. We adopt a heterogeneous model where worker $i$ straggles independently with probability $p_i$ (implying $\mathbb{E}[\mathbb{I}_i]=1-p_i$), which primarily accounts for the computational heterogeneity described in \cite{ref:shiftedexp, ref:MM}.

The global parameters are then updated using the estimated gradient sum $\hat{g}^{(t)}$—rather than the true gradient $g^{(t)}$—following the process in Eq. \eqref{update_param}. Once updated, the new parameters $\boldsymbol{\beta}_{t+1}$ are broadcast to the worker nodes for the next iteration.

\subsection{Quantization Model} 
We adopt the generalized stochastic quantization scheme \cite{ref:qsgd}. Given a vector $\boldsymbol{x} \in \mathbb{R}^l$, worker $i$ applies a quantizer $Q_i(\boldsymbol{x})$ utilizing a bit-width of $z_i$ bits for each individual scalar coordinate. While extending this scheme to layer-wise or bucket-wise granularity is straightforward, we focus on the global formulation here; such extensions typically incur additional communication overhead due to the transmission of multiple scaling factors.

This quantization mechanism preserves the global scale (Euclidean norm) and the sign of each coordinate, while stochastically rounding the magnitude. Specifically, let $s_i = 2^{z_i-1} - 1$ denote the number of quantization levels for the magnitude.  Here, we require a minimum bit-width of $z_i \ge 2$ to ensure at least one valid quantization level for the magnitude (i.e., $s_i \ge 1$) in addition to the sign bit. The quantizer $Q_i(\boldsymbol{x})$ operates coordinate-wise. For the $m$-th coordinate $x_m$, the quantized value is defined as:
\begin{equation} \label{eq:quant_def}
    [Q_i(\boldsymbol{x})]_m = \|\boldsymbol{x}\|_2 \cdot \text{sign}(x_m) \cdot \xi_{i,m},
\end{equation}
where $\|\boldsymbol{x}\|_2$ denotes the $\ell_2$-norm of the vector, and $\xi_{i,m}$ is a random variable representing the quantized normalized magnitude. Letting the normalized magnitude be $u_m = |x_m| / \|\boldsymbol{x}\|_2 \in [0, 1]$, $\xi_{i,m}$ is obtained by performing stochastic rounding on the uniform grid $\{0, 1/s_i, \dots, 1\}$:
\begin{equation} \label{eq:rv_quant}
    \xi_{i,m} = \begin{cases}
        \frac{\lfloor s_i u_m \rfloor + 1}{s_i} & \text{w.p. } s_i u_m - \lfloor s_i u_m \rfloor, \\
        \frac{\lfloor s_i u_m \rfloor}{s_i} & \text{otherwise.}
    \end{cases}
\end{equation}
Consequently, the communication payload consists of the scalar $\|\boldsymbol{x}\|_2$ (e.g., 32 bits) and the quantized coordinates; given the high dimension $l$, the overhead of the real-valued norm is negligible compared to the total bandwidth. 
Crucially, this scheme yields an unbiased estimator of $\boldsymbol{x}$, such that $\mathbb{E}[Q_i(\boldsymbol{x}) \mid \boldsymbol{x}] = \boldsymbol{x}$. 
Furthermore, the variance of the quantization error is bounded by the bit-width. It has been established that there exists a constant $\phi(z_i)$, dependent on the bit-width $z_i$ and dimension $l$, which satisfies:
\begin{equation}
    \mathbb{E}[\|Q_i(\boldsymbol{x}) - \boldsymbol{x}\|_2^2 \mid \boldsymbol{x}] \le \phi(z_i) \|\boldsymbol{x}\|_2^2.
\end{equation}
Typically, $\phi(z_i)$ decays roughly as $O(2^{-2z_i})$, offering a flexible trade-off between communication cost and gradient precision.  Since each normalized coordinate is randomly rounded within a quantization interval of length $1/s_i$, where $s_i=2^{z_i-1}-1$, the coordinate-wise variance is at most $1/(4s_i^2)$, yielding $\phi(z_i)=\frac{l}{4(2^{z_i-1}-1)^2}$. 
In light of this, we note that bucketing serves to mitigate such variance in low-bit $z_i$ regimes \cite{ref:qsgd}.  Note that we employ stochastic rounding that is efficiently implementable via lightweight shared random seeds and is also used in modern low-precision training stacks (e.g., NVIDIA Transformer Engine) to mitigate bias.

\section{Communication-efficient Approximate Gradient Coding} \label{sec:OSGCLight}

Building on the preceding insights, our approach aims to construct a gradient coding scheme that not only minimizes the residual error (i.e., the variance) but also explicitly leverages the unbiased nature of gradient estimators in heterogeneous environments. Crucially, our framework seamlessly integrates communication-efficient quantization directly into the gradient code design. By jointly addressing straggler mitigation and quantization noise, the proposed method enhances convergence stability with rigorous theoretical guarantees. 
Formally, the encoding matrix $A = [a_{i,j}]$, the decoding vector $\mathbf{w} = [w_i]$, and the bit allocation $\mathbf{z} = [z_i]$ are determined by solving the following optimization problem:
\begin{alignat}{2}
	(\textbf{P1}) \quad &\underset{A, \mathbf{w}, \mathbf{z}}{\text{minimize}}	&& 	
	\quad	\mathbb{E}_{S, Q} \big[ \lVert g^{(t)} - \hat{g}^{(t)} \rVert_2^2 \big]\label{opt:p1}\\
	&\text{subject to}													
    && 	\quad \mathbb{E}_{S, Q} \big[ \hat{g}^{(t)} \big] = g^{(t)},\\
   & && 	\quad \sum_{i=1}^{k} z_i = Z_{\text{tot}} \text{ and } z_i \in \mathbb{Z}_{\ge 2},
\end{alignat}
where $Z_{\text{tot}}$ denotes the total bit budget allocated across all workers for quantizing the scalar coordinates of their respective gradient vectors, and each worker needs  at least two bits—allocating one bit for the sign and a minimum of one bit for the magnitude—to ensure the quantization scheme operates correctly.  Here, $\mathbb{E}_{S, Q} [\cdot] \triangleq \mathbb{E} [\cdot|\boldsymbol{\beta}_t]$ denotes the expectation over the stochastic behavior of stragglers and quantizers at iteration $t$, conditioned on the model $\boldsymbol{\beta}_t$.
Notably, the optimized variables remain fixed due to the i.i.d. nature of stragglers across iterations, requiring updates only when system statistics shift.

However, a direct solution to (\textbf{P1}) is intractable as it depends on the unknown true gradients $g_j^{(t)}$. To derive a realizable code, we adopt a standard boundedness assumption consistent with the literature \cite{ref:SGC, ref:HAGC, ref:SGC2, ref:FL}.

\begin{assumption} \label{assumption:const}
There exists a constant $C > 0$ such that the squared $\ell_2$-norm of the partial gradient is bounded:
\begin{equation}
 \lVert \nabla L(\mathcal{D}_j, \boldsymbol{\beta}_t) \rVert_2^2 = \lVert g_j^{(t)}\rVert_2^2 \leq C, \quad \forall j \in [1:n].
\end{equation}
\end{assumption}

Under Assumption~\ref{assumption:const}\footnote{ Assumption~\ref{assumption:const} can be relaxed to an affine growth condition by replacing $C$ with $\mathsf{a}\|\nabla L(\boldsymbol{\beta}_t)\|_2^2+\mathsf{b}$ in the error bounds;  the subsequent derivation of the proposed design remains unchanged.}, we derive a tractable upper bound for the objective function.

\begin{lemma} \label{lem1_quant}
Suppose Assumption~\ref{assumption:const} holds and the estimator $\hat{g}^{(t)}$ is unbiased. Then, the error is bounded by:
\begin{alignat}{2} 
	& &&\mathbb{E}_{S, Q} \big[ \lVert g^{(t)} - \hat{g}^{(t)} \rVert_2^2 \big]  \nonumber \\
    & &&  \leq  C \sum_{i=1}^k (1-p_i)\big\{ p_i+ \phi(z_i) \big\} \cdot w_i^2 \bigg(\sum_{j=1}^n a_{i,j}\bigg)^2,
\end{alignat}
 where $\phi(z) \triangleq \frac{l}{4 \cdot (2^{z-1}-1)^2}$ arises from the quantization error, and $n$ denotes the number of data partitions.
\end{lemma}
The proof is provided in Appendix~\ref{apdx:lem1}. 
Next, we analyze the unbiasedness constraint. Since the quantizers are unbiased, the expected estimator is given by:
\begin{equation} \label{eq:exp_g1_quant}
 \mathbb{E}_{S, Q} \big[ \hat{g}^{(t)} \big]
 = \sum_{j=1}^n g_j^{(t)} \cdot \sum_{i=1}^k (1-p_i) w_i a_{i,j}.
\end{equation}
To ensure $\mathbb{E}_{S, Q} [\hat{g}^{(t)}] = \sum_{j=1}^n g_j^{(t)}$ for any realization of gradients, the coding coefficients must satisfy the following partition-wise condition:
\begin{equation} \label{eq:exp_g2_quant}
 \sum_{i=1}^k (1-p_i) w_i a_{i,j} = 1, \quad \forall j \in [1:n].
\end{equation}
 
Therefore, by defining $\tilde{w}_i \triangleq (1-p_i) w_i$ for simplicity and using Lemma \ref{lem1_quant}, (\textbf{P1}) can be reformulated as follows.
\begin{alignat}{2}
	(\textbf{P2}) \quad &\underset{A, \mathbf{w}, \mathbf{z}}{\text{minimize}}	&&
	\quad	\sum_{i=1}^k \frac{p_i +\phi(z_i)}{1-p_i} \tilde{w}_i^2 \bigg( \sum_{j=1}^n a_{i,j} \bigg)^2 \label{opt:p2}\\
	&\text{subject to}													&& 	\quad
	\sum_{i=1}^k \tilde{w}_i a_{i,j} = 1, ~ \forall j \in [1:n],  \\
   & && 	\quad \sum_{i=1}^{k} z_i = Z_{\text{tot}} \text{ and } z_i \in \mathbb{Z}_{\ge 2}.
\end{alignat}
The joint optimization (\textbf{P2}) is inherently intractable as it constitutes a Mixed-Integer Non-Linear Programming (MINLP) problem due to the discrete integer constraints for quantization. To address this intractability, we decouple the problem: we first derive the closed-form optimal code structure for a fixed bit allocation, then optimize the bit allocation. As we will demonstrate, by analytically resolving the inner dependency between the code structure and quantization, this sequential approach preserves the joint optimality.

\subsection{Optimal Structure of Quantized Gradient Code}
To mitigate the intractability of the joint design problem, we first aim to identify the optimal structure of the quantized gradient code for a given bit allocation $\mathbf{z}$. 
Let $c_i \triangleq \frac{p_i +\phi(z_i)}{1-p_i}$, 
 which is a determined value for a given bit allocation.
Then, the problem (\textbf{P2}) is reformulated as (\textbf{c-SP1}):
\begin{alignat}{2}
	(\textbf{c-SP1}) \quad &\underset{A, \mathbf{w}}{\text{minimize}}	&&
	\quad	\sum_{i=1}^k c_i \tilde{w}_i^2 \bigg( \sum_{j=1}^n a_{i,j} \bigg)^2 \label{opt:p2q}\\
	&\text{subject to}													&& 	\quad
	\sum_{i=1}^k \tilde{w}_i a_{i,j} = 1, ~ \forall j \in [1:n].
\end{alignat}
While Problem (\textbf{c-SP1}) is non-convex due to the bilinear terms $\tilde{w}_i a_{i,j}$, it admits a convex reformulation. By introducing the auxiliary variable $\alpha_i^j \triangleq \tilde{w}_i a_{i,j}$, which represents the \emph{effective weight} of worker $i$ for partition $j$, we transform (\textbf{c-SP1}) into the following convex problem (\textbf{c-SP2}):

\begin{alignat}{2}
	(\textbf{c-SP2}) \quad &\underset{\boldsymbol{\alpha}}{\text{minimize}}	&& 	
	\quad	\sum_{i=1}^k c_i \bigg( \sum_{j=1}^n \alpha_i^j \bigg)^2 \label{opt:p3q}\\
	&\text{subject to}													&& 	\quad
	\sum_{i=1}^k \alpha_i^j = 1, \quad \forall j \in [1:n].
\end{alignat}
Here, the matrix $\boldsymbol{\alpha}$ explicitly encodes the data partition distribution structure. Once $\boldsymbol{\alpha}$ is determined, the physical coefficients are recovered as $a_{i,j} = \alpha_i^j / \tilde{w}_i$ and $w_i = \tilde{w}_i / (1-p_i)$ for any non-zero $\tilde{w}_i$, which can be randomly generated.

While Problem (\textbf{c-SP2}) is solvable via standard tools like CVX \cite{ref:CVX}, we derive its closed-form optimal structure to gain theoretical insights. 
	\begin{theorem} \label{thm:opt_str}
		Given $\{p_i\}$ and $\{z_i\}$, the optimal structure of optimization problem (\textbf{c-SP2}) satisfies the conditions below:
		\begin{equation}
			 \sum_{j=1}^n \alpha_i^j = Y_i, \forall i\in[1:k], \text{ and }
			 \sum_{i=1}^k \alpha_i^j = 1, \forall j\in[1:n],
		\end{equation}
		where $Y_i=c_i^{-1}\cdot \frac{n}{\sum_{j=1}^k c_j^{-1}}$ and $c_i = \frac{p_i +\phi(z_i)}{1-p_i}$.
	\end{theorem}
		The proof is provided in Appendix \ref{apdx:thm1}. 
Note that the constraint $\sum_{i=1}^k \alpha_i^j = 1$ ensures the unbiasedness of the gradient estimator. The optimal values of $\boldsymbol{\alpha}$ are governed by $Y_i$, which captures the interplay between straggler probabilities and bit allocations. Consequently, the resulting gradient code is explicitly shaped by the heterogeneous characteristics (straggling behavior and bit budgets) of the workers.

Based on this optimal structure, we provide a closed-form bound on the estimator's performance.
\begin{lemma} \label{lem:ub_err}
For any gradient coding scheme satisfying the optimal structure, the residual error is bounded by:
\begin{equation}
 \mathbb{E}_{S, Q} \big[ \lVert g^{(t)} - \hat{g}^{(t)}\rVert_2^2 \big]
 \leq \frac{n^2 C}{\sum_{i=1}^k c_i^{-1}}.
\end{equation}
\end{lemma}
The proof is provided in Appendix~\ref{apdx:lem_ub_err}.

Although multiple configurations may satisfy the optimal structure in Theorem~\ref{thm:opt_str}, we present a closed-form sparse construction that achieves low computation load. 
Since $c_i>0$, we have $Y_i>0$ and $\sum_{i=1}^k Y_i=n$. 
Define the cumulative sums $R_0=0$ and $R_i=\sum_{m=1}^i Y_m$ for $i\in[1:k]$. 
We represent the target mass $Y_i$ of worker $i$ by a contiguous allocation segment $[R_{i-1},R_i]$ on the data-partition axis $[0,n]$.

Each data partition $\mathcal{D}_j$ is associated with the unit segment $[j-1,j]$. We then define $\alpha_i^j$ as the overlap length between worker $i$'s allocation segment and the unit segment of $\mathcal{D}_j$:
\begin{equation}
    \alpha_i^j = \left[\min\{R_i,j\} - \max\{R_{i-1},j-1\} \right]_+,
\end{equation}
where $[x]_+=\max\{x,0\}$ and this construction immediately gives $\alpha_i^j\ge0$. Moreover, since the length of worker $i$'s allocation segment is $Y_i$, we have $\sum_{j=1}^n\alpha_i^j=Y_i$; and since each unit segment $[j-1,j]$ is fully covered by the allocation segments, we have $\sum_{i=1}^k\alpha_i^j=1$. Thus, the construction preserves the optimal structure.

Nonzero entries arise only when an allocation segment overlaps a unit data-partition segment. Hence, the number of nonzero entries is at most $n+k-1$, and the computation load satisfies $d\le 1+\frac{k-1}{n}$. Since $k\le n$ in typical distributed learning settings, this gives $d\le2$, achieving low redundancy while maintaining straggler robustness.

\subsection{Optimal Bit Allocation under the Optimal Structure}
\label{subsec:proposed_algo}
Building upon the optimal structure of the gradient code in the preceding subsection, we now address the bit allocation optimization, ensuring that the optimality of the joint design remains intact.  
We consider $k$ workers with a total budget $Z_{\text{tot}}$, where $z_i$ denotes the bits per coordinate. By targeting $z_i \ll 32$ (standard float precision), our scheme achieves significant communication compression.

 To ensure the valid operation of the quantization scheme, each worker requires at least two bits per coordinate. 
We formulate the problem in terms of integers $z_i \in \mathbb{Z}_{\ge 2}$ subject to $\sum_{i=1}^{k} z_i = Z_{\text{tot}}$.
Let $r_i \triangleq z_i - 2 \in \mathbb{Z}_{\ge 0}$ denote the \emph{residual bits} assigned to worker $i$ beyond the mandatory allocation. 
Defining the residual budget as $Z_{\text{res}} \triangleq Z_{\text{tot}} - 2k$, the optimization problem is equivalent to determining the residual allocation vector $\mathbf{r} = [r_1, \dots, r_k]^T$:
\begin{alignat}{2} 
	(\textbf{b-SP1}) \quad &\underset{\mathbf{r}}{\text{maximize}}	&& 	
	\quad F(\mathbf{r}) \triangleq	\sum_{i=1}^{k} h_i(r_i) \label{opt:p1b}\\
	&\text{subject to}													&& 	\quad
  \sum_{i=1}^{k} r_i = Z_{\text{res}} \text{ and } r_i\in \mathbb{Z}_{\ge 0}, 
\end{alignat}
where $h_i(r_i)=\frac{1-p_i}{p_i + \phi(r_i+2)}$ and $\phi(z) = \frac{l}{4 \cdot (2^{z-1}-1)^2}$. 

The following proposition establishes that the bit allocation formulation in (\textbf{b-SP1}) fully preserves the optimality of the original joint problem (\textbf{P2}). It proves that maximizing the utility sum under Theorem \ref{thm:opt_str} is mathematically equivalent to minimizing the original residual error.
\begin{proposition} \label{prop:bitopt}
    Optimizing $\mathbf{r}$ based on the derived optimal code structure yields the global optimum of (\textbf{P2}), ensuring that the reduction to (\textbf{b-SP1}) incurs no loss of optimality.
\end{proposition}
The proof is provided in Appendix~\ref{apdx:prop_bitopt}.

\subsubsection{Optimal Dynamic Programming Algorithm}
We first describe an optimal solution using Dynamic Programming (DP). Define $V[i, r]$ as the maximum utility achievable using a subset of workers $\{1, \dots, i\}$ with a cumulative residual budget $r$.  Since the objective function is a sum of separable functions $h_i(r_i), \forall i$, the problem satisfies the optimal substructure property, leading to the following Bellman equation governing the state transitions: 
\begin{equation}
  V[i, r] = \max_{0 \le \mathsf{a} \le r} \big( V[i-1, r-\mathsf{a}] + h_i(\mathsf{a}) \big),
  \label{eq:dp_rec}
\end{equation}
with boundary conditions $V[0, 0] = 0$ and $V[0, r] = -\infty$.

\begin{lemma} \label{lem:dp_opt}
The recurrence in \eqref{eq:dp_rec} guarantees finding the global optimum $F^\star$ of the integer optimization problem (\textbf{b-SP1}), i.e., $F^\star = V[k, Z_{\text{res}}]$.
\end{lemma}
The proof is provided in Appendix \ref{apdx:lem_dpopt}. It is worth noting that the joint global optimum for (\textbf{P2}) can be attained through the combination of the derived optimal structure and DP. 
However, while DP yields the global optimum, its computational complexity is $O(k Z_{\text{res}}^2)$, which becomes prohibitive for large-scale learning models where $Z_{\text{tot}}$ is large. 

\subsubsection{Proposed Low-Complexity Algorithm}
 To address the computational overhead of optimal DP algorithm, we propose a low-complexity algorithm that efficiently finds a near-optimal solution. 
The design of our proposed algorithm is theoretically grounded in two key structural properties of the utility function $h_i(r)$.

First, for any fixed bit allocation $r \ge 0$, the utility contribution $h_i(r)$ is strictly decreasing with respect to the straggling parameter $p_i$. That is, $\frac{\partial h_i(r)}{\partial p_i} = - \frac{\phi (r+2) + 1}{(\phi (r+2) + p_i)^2} < 0$, since $\phi (r+2) > 0$. This monotonicity implies that, all else being equal, workers with higher reliability (smaller $p_i$) inherently contribute more to the objective function. This structural insight provides a critical operational advantage: it justifies pruning the exponential search space of worker subsets down to a linear number of candidates. Specifically, it supports the greedy sorting strategy (Line 1 of Algorithm \ref{alg:proposedba}), where we prioritize allocating resources to the nested subsets of the most reliable workers. We denote this sorted set as $S_\kappa \triangleq \{i_1, \dots, i_\kappa\}$ with $p_{i_1} \le \dots \le p_{i_\kappa}$, and the algorithm iterates through these Top-$\kappa$ candidates.

Second, the function $h_i(r)$ exhibits a sigmoidal geometry, which necessitates a regime-adaptive allocation strategy based on the marginal gain $\Delta_i(r)\triangleq h_i(r+1)-h_i(r)$.
\begin{itemize}
    \item \textbf{Convex Regime:} 
    In the low-budget convex region, marginal gains increase with $r$. Thus, consolidating bits on a few workers yields higher utility than spreading them. Our algorithm explicitly realizes this concentration by iterating through small $\kappa$ (e.g., $\kappa=1,2$).
    \item \textbf{Concave Regime:} 
    Conversely, in the high-budget concave region, marginal gains diminish due to saturation. Consequently, total utility is maximized by distributing bits to harvest higher marginal gains from multiple workers, a strategy naturally approximated by the Lagrangian relaxation which inherently seeks to equalize the derivatives across all active workers.
    \item \textbf{Transition Regime:} Near the inflection point, equal bit allocation acts as a stable baseline to mitigate the Lagrangian solution's instability.
\end{itemize}

\begin{algorithm}[t]
\caption{Proposed Low-Complexity Bit Allocation}
\label{alg:proposedba}
\begin{algorithmic}[1]
\STATE Sort workers $p_{i_1} \le \dots \le p_{i_k}$ and set $\mathcal{F}_{\max} \leftarrow -\infty$.
\FOR{$\kappa = 1$ to $\min(k, Z_{\text{res}})$}
    \STATE Compute $\mathbf{r}^{\text{LAG}}$ and $\mathbf{r}^{\text{EQ}}$ on top-$\kappa$ workers.
    \STATE $\mathbf{r}^{(\kappa)} \leftarrow \arg\max_{\mathbf{r} \in \{\mathbf{r}^{\text{LAG}}, \mathbf{r}^{\text{EQ}}\}} F(\mathbf{r})$.
    \STATE Refine $\mathbf{r}^{(\kappa)}$ to $\tilde{\mathbf{r}}^{(\kappa)}$ via iterative 1-bit swaps.
    \STATE Update: if $F(\tilde{\mathbf{r}}^{(\kappa)}) > \mathcal{F}_{\max}$ then $\mathbf{r}^\star \leftarrow \tilde{\mathbf{r}}^{(\kappa)}$.
\ENDFOR
\end{algorithmic}
\end{algorithm}

Accordingly, our algorithm iterates through the active set size $\kappa$ to implicitly handle the non-convexity. 
Let $\kappa_{\max} = \min(k, Z_{\text{res}})$. The algorithm iterates $\kappa$ from $1$ to $\kappa_{\max}$, performing the following steps for each active set $S_\kappa$: 
\begin{enumerate}
    \item \textbf{Lagrangian Relaxation:} We obtain $\mathbf{r}^{\text{LAG}}(\kappa)$ by solving the continuous relaxation via Lagrange multipliers and rounding the result.
  
    \item \textbf{Equal Allocation:} We compute $\mathbf{r}^{\text{EQ}}(\kappa)$ by distributing bits uniformly among $S_\kappa$ (differing by at most 1 bit) to ensure robustness against non-convexity.
  
  \item \textbf{Hybrid Selection:} We select the better candidate for the given $\kappa$: 
    $\mathbf{r}^{(\kappa)} = \arg\max_{\mathbf{r} \in \{ \mathbf{r}^{\text{LAG}}(\kappa), \mathbf{r}^{\text{EQ}}(\kappa) \}} F(\mathbf{r}).$
  
  \item \textbf{One-Bit Local Search (Refinement):} 
  We refine $\mathbf{r}^{(\kappa)}$ by iteratively applying 1-bit swaps. A swap from worker $j$ to $i$ is accepted if the gain is positive:
  $[h_i(r_i+1) - h_i(r_i)] + [h_j(r_j-1) - h_j(r_j)] > 0.$ 
  This process converges to a local optimum $\tilde{\mathbf{r}}^{(\kappa)}$.
\end{enumerate}

Finally, the algorithm selects the global best configuration $\kappa^\star = \arg\max_\kappa F(\tilde{\mathbf{r}}^{(\kappa)})$. The computational complexity is dominated by the sorting and per-$\kappa$ operations, resulting in an empirical complexity of approximately $O(k^2)$, which offers a significant reduction compared to DP while maintaining near-optimal performance, as we will demonstrate in the experimental section. The details are in Algorithm \ref{alg:proposedba}.


    \section{Theoretical Analysis} \label{sec:conv}
	In this section, we provide a rigorous convergence analysis of the proposed optimally structured gradient coding scheme for various classes of loss functions. To ensure analytical tractability, we adopt the following standard assumptions regarding the objective function:
	
	\begin{assumption} \label{assumption:conv} ($\lambda$-strongly convexity)
		The loss function $L$ is $\lambda$-strongly convex if for all $\boldsymbol{\beta},\boldsymbol{\beta}'\in \mathbb{R}^l$,
		\begin{equation}
		L(\boldsymbol{\beta}) \,\geq\, L(\boldsymbol{\beta}') + \langle \nabla L(\boldsymbol{\beta}') , \boldsymbol{\beta} - \boldsymbol{\beta}' \rangle 
		\;+\; \frac{\lambda}{2}\,\bigl\|\boldsymbol{\beta} - \boldsymbol{\beta}'\bigr\|^2_2.
		\end{equation}
	\end{assumption}
	
	\begin{assumption} \label{assumption:smooth} ($\mu$-smoothness)
		The loss function $L$ is $\mu$-smooth if for all $\boldsymbol{\beta},\boldsymbol{\beta}'\in \mathbb{R}^l$, $\mu\geq0$,
		\begin{equation}
		L(\boldsymbol{\beta}) \,\leq\, L(\boldsymbol{\beta}') + \langle \nabla L(\boldsymbol{\beta}') , \boldsymbol{\beta} - \boldsymbol{\beta}' \rangle 
		\;+\; \frac{\mu}{2}\,\bigl\|\boldsymbol{\beta} - \boldsymbol{\beta}'\bigr\|^2_2.
		\end{equation}
	\end{assumption}  
	Here, $\langle \cdot, \cdot \rangle$ denotes the inner product operation.
	We remark that \textbf{Assumptions \ref{assumption:conv}} and \textbf{\ref{assumption:smooth}} are standard conditions satisfied by a wide range of popular learning models, including logistic regression and softmax classifiers.
	
	We begin by establishing the convergence guarantees for $\lambda$-strongly convex loss functions in the following theorem.
	\begin{theorem} \label{thm:sc} Suppose the loss function $L$ satisfies \textbf{Assumptions \ref{assumption:const}} and \textbf{\ref{assumption:conv}}. Then, by setting $\gamma_t = 1/(\lambda t)$,  which decays with the increasing training iteration $t$, it holds for any optimally structured gradient codes that 
		\begin{equation}
		\mathbb{E} [\lVert \boldsymbol{\beta}_T - \boldsymbol{\beta}^* \rVert^2_2]\leq \frac{4 n^2 C}{\lambda^2 T}  \bigg(1+\frac{1}{\sum_{i=1}^k c^{-1}_i} \bigg),
		\end{equation}
		where $c^{-1}_i=\frac{1-p_i}{p_i+\phi(z_i)}$.
	\end{theorem}
		The proof is provided in Appendix \ref{apdx:thm2}. 
	 Theorem \ref{thm:sc} establishes that for strongly convex loss functions with a decaying learning rate $\gamma_t = \frac{1}{\lambda t}$ that decays as the training iteration $t$ increases, the proposed method achieves a convergence rate of $O(1/T)$, matching that of classical SGD. Notably, the derived error bound explicitly incorporates both straggler heterogeneity and bit allocation strategies via the scaling factor $\left( 1 + \frac{1}{\sum_{i=1}^k c_i^{-1}} \right)$, ensuring robust convergence even under non-uniform straggling conditions and varying quantization budgets.  A critical advantage of our approach is that the convergence rate is theoretically independent of the computation load $d$, as it is determined solely by the heterogeneous capacities. Consequently, minimal data replication is sufficient. This stands in sharp contrast to SGC \cite{ref:SGC}, where convergence guarantees are contingent upon the minimum computation load $\min_{i \in [1:k]} d_i$, thereby underscoring the superior robustness of our framework.
	
	Under \textbf{Assumption \ref{assumption:smooth}}, we extend the convergence analysis to $\mu$-smooth, possibly non-convex losses.
	\begin{theorem}\label{thm:s}
		Suppose the loss function $L$ satisfies \textbf{Assumptions \ref{assumption:const}} and \textbf{\ref{assumption:smooth}}. Then, it holds for any optimally structured gradient codes: 
		
		By setting $\gamma_t=\gamma = 1/(T+1)^{1/2}$,  which remains fixed regardless of the increasing training iteration $t$, 
		\begin{align}
			\frac{1}{T+1} \sum_{t=0}^T &\mathbb{E} [\lVert g^{(t)} \rVert^2_2] \leq \frac{L(\boldsymbol{\beta}_{0}) - L(\boldsymbol{\beta}^*)}{(T+1)^{1/2}}  \nonumber \\
			&+\frac{1}{(T+1)^{1/2}}\frac{\mu   n^2 C}{2} \cdot  \bigg(1+\frac{1}{\sum_{i=1}^k c^{-1}_i} \bigg).
		\end{align}
		
		By setting $\gamma_t={1}/{(t+1)^{1/2}}$,  which decays with the increasing training iteration $t$, 
		\begin{align}
			\frac{1}{T+1} &\sum_{t=0}^T \mathbb{E} [\lVert g^{(t)} \rVert^2_2] \leq \frac{L(\boldsymbol{\beta}_{0}) - L(\boldsymbol{\beta}^*)}{(T+1)^{1/2}}  \nonumber \\
			&+ \frac{\mu   n^2 C (1+\log(T+1)^{1/2})\bigg(1+\frac{1}{\sum_{i=1}^k c^{-1}_i} \bigg)}{(T+1)^{1/2}}.
		\end{align}
        
        Thus, they hold the following limit: 
        \begin{equation}
            \lim_{T\rightarrow\infty} \frac{1}{T+1} \sum_{t=0}^T \mathbb{E} [\lVert g^{(t)} \rVert^2_2] = 0.
        \end{equation}
		
	\end{theorem}
		The proof is provided in Appendix \ref{apdx:thm3}. 
    Theorem \ref{thm:s} establishes that for $\mu$-smooth loss functions, the proposed gradient coding scheme guarantees convergence to a stationary point.  Specifically, under both constant ($\gamma_t = (T+1)^{-1/2}$) and decaying ($\gamma_t = (t+1)^{-1/2}$) learning rate schedules with respect to the increasing training iteration $t$, the average squared gradient norm asymptotically approaches zero. This result validates the algorithm's effectiveness in general non-convex settings\footnote{ For non-smooth losses, the argument follows standard subgradient-based convergence results under unbiasedness and bounded variance, as discussed in Appendix C.4.3 of our previous work \cite{ref:HAGC}.}.

	
	\section{Discussions} \label{sec:dis}
	\subsection{Practical Example of Probabilistic Straggler Modeling}
    Let $\tau_{th}$ represent the response time deadline for each training iteration. A worker node $i \in [1:k]$ is identified as a straggler if its total delay $\tau_i$, encompassing both local gradient computation and communication, exceeds this threshold (i.e., $\tau_i > \tau_{th}$). Consequently, the straggler probability for worker $i$ is defined as:
    \begin{equation}
    p_i = \text{Pr}(\tau_i > \tau_{th}).
    \end{equation}
    The statistical modeling of stragglers is contingent upon the system's primary bottleneck. In communication-limited environments, straggler behavior is typically described by an exponential distribution \cite{ref:Buyukates}, whereas in computation-limited settings, latency is often characterized by a shifted-exponential distribution \cite{ref:MM, ref:shiftedexp}. Beyond these parametric models, practical estimation techniques like empirical maximum likelihood estimation (MLE) can also be employed to infer straggler probabilities from historical data.
    
    Intuitively, the likelihood of a worker becoming a straggler depends not only on its stochastic latency—governed by computational and communication capacities—but also on the threshold $\tau_{th}$. While a tighter threshold accelerates model updates, it inevitably increases the straggler probability, revealing a fundamental trade-off between iteration speed and straggler mitigation. Our empirical results demonstrate that the proposed scheme effectively balances this trade-off.

    \subsection{Practical Scaling for Error Balance}
    Depending on the distributed learning environment, the magnitude of the quantization error variance may disproportionately dominate or diminish relative to the straggler probabilities. To mitigate this potential disparity and prevent the optimization from being biased toward a single error source, a scaling factor $\eta$ can be introduced as a practical modification. Under this practical scheme, $c_i$ is adapted as:    
    \begin{equation} 
    	c_i = \frac{p_i}{1-p_i} + \eta \cdot \frac{\phi(z_i)}{1-p_i}.
    \end{equation}
    For instance, to align the magnitude of the quantization error variance with the average straggler probability for balanced optimization, a practical configuration is to set $\eta = ( \frac{1}{k}\sum_{i=1}^k p_i ) \cdot 4(2^{Z_{\text{tot}}/k-1}-1)^2/l$. Beyond this baseline, $\eta$ can serve as a tunable hyperparameter to trade off between communication efficiency and training stability depending on the system requirements.

\subsection{Sparsity of Optimally Structured Gradient Coding}

The sparsity of $\boldsymbol{\alpha}$ determines the number of non-zero encoding coefficients in $A$, where each non-zero entry $\alpha_i^j$ corresponds to an assignment between worker $i$ and data partition $\mathcal{D}_j$. 
Thus, reducing the number of non-zero entries directly reduces the computation load. 
Following the graph-based sparse construction in Appendix C.3 of our previous work \cite{ref:HAGC}, we view $\boldsymbol{\alpha}$ as a weighted bipartite graph whose row and column vertices represent workers and data partitions, respectively. 
The goal is to remove unnecessary edges while preserving the optimal row-wise constraints $\sum_{j=1}^n\alpha_i^j=Y_i$ and the unbiasedness constraints $\sum_{i=1}^k\alpha_i^j=1$.

To this end, the graph can be decomposed into disjoint connected subgraphs. 
If the resulting decomposition consists of $o_{\max}$ subgraphs $\{\mathcal{S}_l\}_{l=1}^{o_{\max}}$ with row and column sets $\mathcal{R}_l$ and $\mathcal{C}_l$, respectively, each subgraph can be realized with the minimum number of edges required for connectivity, i.e., $|\mathcal{R}_l|+|\mathcal{C}_l|-1$. 
Accordingly, the resulting computation load is
    \begin{equation}
        d = \frac{\sum_{l=1}^{o_{\textsf{max}}} (|\mathcal{R}_l| + |\mathcal{C}_l| - 1)}{n}.
    \end{equation}
The detailed recursive partitioning and construction algorithm is provided in Appendix C.3 of our previous work \cite{ref:HAGC}.

    \subsection{Theoretical Justification for Full-Batch Formulation} \label{sec:sgd}
    Although the optimization problem \textbf{(P1)} is formulated based on full-batch gradients, our experiments employ mini-batch SGD. To justify this, we denote the random variables associated with the system and training process as follows: $S$ represents the straggler realizations, $Q$ denotes the stochastic quantization noise, and $\Xi$ represents the randomness from mini-batch sampling.
    
    Let $\bar{g}^{(t)}$ denote the aggregated gradient estimator constructed from mini-batch stochastic gradients (as used in experiments). Correspondingly, let $\hat{g}^{(t)}$ represent the estimator derived using full-batch gradients under the identical straggler realization $S$ and quantization parameters.
    Crucially, since the mini-batch gradient is an unbiased estimator of the full-batch gradient, we have $\mathbb{E}_{\Xi}[\bar{g}^{(t)} \mid S, Q] = \hat{g}^{(t)}$.
    Assuming that the sampling randomness $\Xi$ is independent of the system-induced randomness $(S, Q)$, the total expected error $\mathbb{E}_{S,Q,\Xi}\!\left[\left\|g^{(t)}-\bar{g}^{(t)}\right\|_2^2\right]$ can be decomposed by taking the expectation over their joint distribution $(S, Q, \Xi)$:
    \begin{equation}
        \underbrace{\mathbb{E}_{S,Q}\!\left[\left\|g^{(t)}-\hat{g}^{(t)}\right\|_2^2\right]}_{\text{(I) System-induced error}}+\underbrace{\mathbb{E}_{S,Q,\Xi}\!\left[\left\|\hat{g}^{(t)}-\bar{g}^{(t)}\right\|_2^2\right]}_{\text{(II) Sampling error}}.
    \end{equation}
    Term (I) corresponds exactly to the objective function of \textbf{(P1)}, capturing distortions solely induced by heterogeneous stragglers and quantization.
    In contrast, Term (II) accounts for the additional variance arising from mini-batch sampling. Explicitly optimizing this sampling variance typically favors denser weighting schemes (i.e., full replication $d=k$) to average out noise, as discussed in \cite{ref:HAGC}, which directly conflicts with our goal of communication- and computation-efficient coding.
    Therefore, our framework deliberately focuses on minimizing Term (I), while Term (II) is effectively managed through standard SGD mechanisms (e.g., mini-batch size adjustments and learning-rate schedules), as validated empirically in Section \ref{sec:exp}.

\subsection{Extension to Adaptive Optimizer} \label{sec:adam}
    
    While strictly unbiased gradient estimation favors GD/SGD-based optimization, it creates a detrimental trade-off for adaptive optimizers (e.g., Adam). Specifically, enforcing zero bias compels the estimation error to manifest entirely as variance. Since the second-moment estimator reflects both squared bias and variance, this inflated variance systematically overestimates the gradient scale, leading to excessive step-size shrinkage. To address this bias–variance tradeoff without incurring additional communication overhead, we propose a two-track decoding strategy. This approach maintains the original unbiased decoding for the first moment while employing a separate, variance-reduced decoding vector for the second moment. This effectively decouples the estimation objectives, allowing the master node to apply two distinct decoding vectors to the same received signals.
    
    Since the encoding structure is fixed by our scheme, we focus on optimizing the second-moment decoder $v$ to minimize a weighted combination of squared bias and variance:
    \begin{align} \label{eq:twodec}
    \min_{v} &  ~\Lambda  \left[\sum_j \left(1- \sum_i (1-p_i)v_i a_{i,j}\right)\right]^2  \nonumber \\ 
    & + \sum_i (1-p_i) \{p_i+\phi(z_i)\}\cdot  v_i^2 \left(\sum_j a_{i,j}\right)^2,
    \end{align}
    where $\Lambda$ controls the emphasis on bias reduction. The closed-form optimal solution is derived as $v_i^*=\frac{n \Lambda}{(p_i+\phi(z_i)) (\sum_{j=1}^n a_{i,j})(1+\Lambda \sum_{m=1}^k c^{-1}_m)}$.  By utilizing the decoding coefficients $w^*$ for the first moment and $v^*$ for the second, this method significantly improves convergence stability in heterogeneous environments with zero additional transmission cost (see Proposition \ref{prop:twodec} and Section \ref{sec:exp}).
    \begin{proposition}     \label{prop:twodec}
    The decoder $v^*$ minimizes the second-moment decoding error than any unbiased decoder, which reduces the variance of the adaptive denominator and stabilizes adaptive optimization. 
    \end{proposition}
    The proof is provided in Appendix \ref{apdx:twodec}.

	\section{Experiments} \label{sec:exp}
	    
    This section validates the efficacy of our proposed communication-efficient optimally structured gradient coding scheme for mitigating stragglers in distributed learning environments. We conduct numerical evaluations using the extensive COCO dataset \cite{ref:coco}, which contains 118,287 training and 5,000 validation images across 80 object categories. Recognized as a rigorous benchmark for object detection, instance segmentation, and image captioning, the COCO dataset serves as a comprehensive testbed for our approach. Experiments are performed using  the MobileNetV3 architecture, a state-of-the-art model widely adopted for efficient object detection on the COCO dataset, with a learning rate of $\gamma_t = 0.05$.
    Let $\tau_{th}$ represent the response time deadline for a training iteration. A worker node $i \in [1:k]$ is identified as a straggler if its total delay $\tau_i$, comprising local gradient computation and communication time, exceeds this threshold (i.e., $\tau_i > \tau_{th}$). Consequently, the straggler probability for worker node $i$ is determined by
    \begin{equation} 
    p_i = Pr(\tau_i > \tau_{th})=e^{-\psi_i (\tau_{th} - 1)},
    \end{equation} 
	where $\psi_i$ represents the straggling parameter of the worker node $i$ and $\tau_{th} \geq 1$ \cite{ref:MM, ref:shiftedexp}. 
	In these experiments, the straggling parameter $\psi_i$ is drawn from a uniform distribution defined as $\text{Uniform}(\psi_\textsf{min}, \psi_\textsf{max})$. Unless otherwise specified, we employ the following default parameter settings: $k=10$, $\psi_\textsf{min}=0.1$, $\psi_\textsf{max}=2$, and $\tau_{th}=1.1$. To ensure statistical reliability, all results represent the average across 50 independent simulation runs.
	\begin{figure}[t]
		\centering
		\begin{subfigure}[b]{0.315\textwidth} 
			\includegraphics[width=\linewidth]{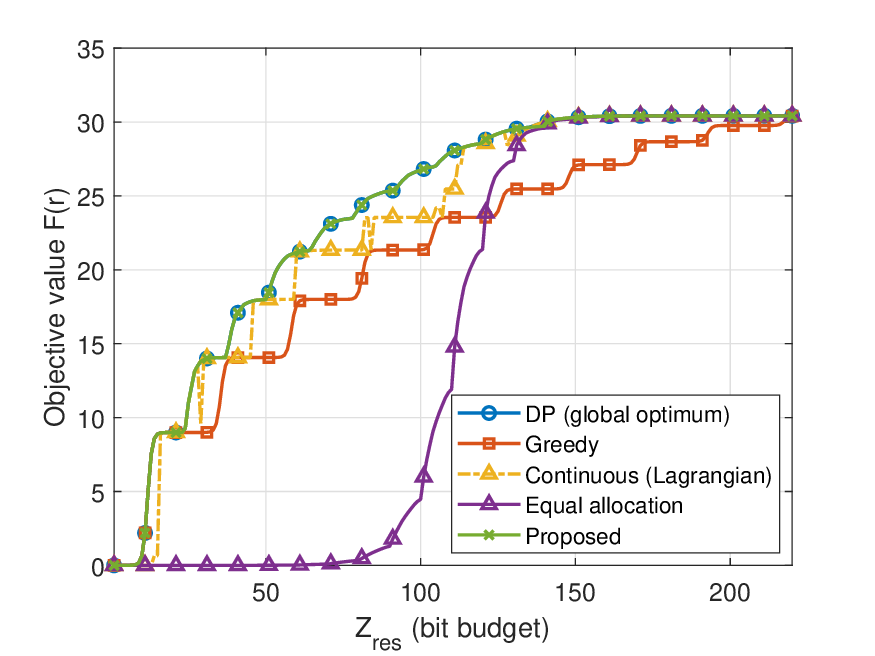}
			\caption{}
			\label{fig:ba1}
		\end{subfigure}
		\begin{subfigure}[b]{0.315\textwidth}
			\includegraphics[width=\linewidth]{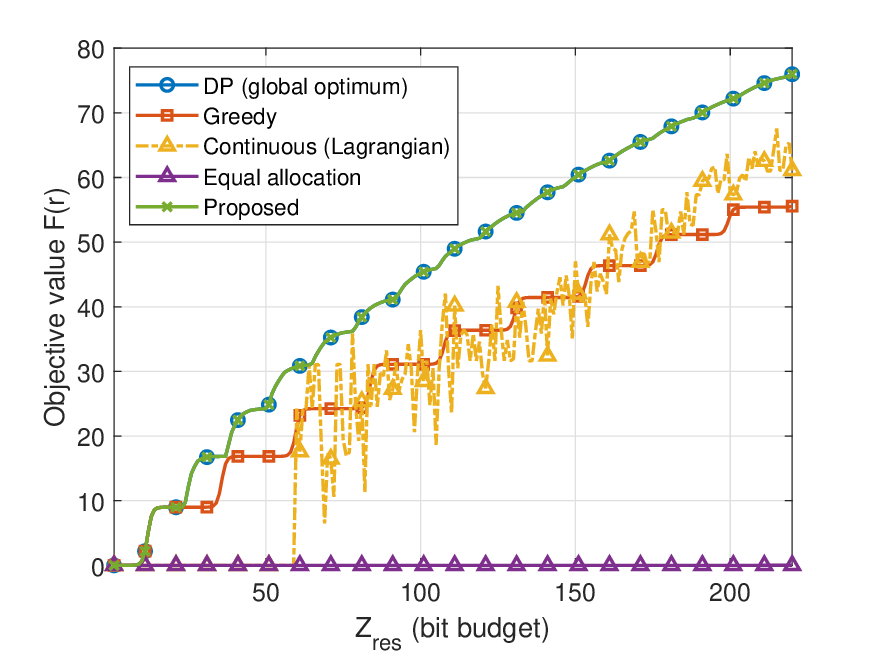}
			\caption{}
			\label{fig:ba2}
		\end{subfigure}
		\caption{The objective value $F(\mathbf{r})$ for (\textbf{b-SP1}) with respect to the bit budget $Z_\text{res}$: (a) $k=10$ and (b) $k=50$.}
		\label{fig:ba}
		\vspace{-10pt}
	\end{figure}
	\begin{figure*}[t]
		\centering
		\begin{subfigure}[b]{0.315\textwidth} 
			\includegraphics[width=\linewidth]{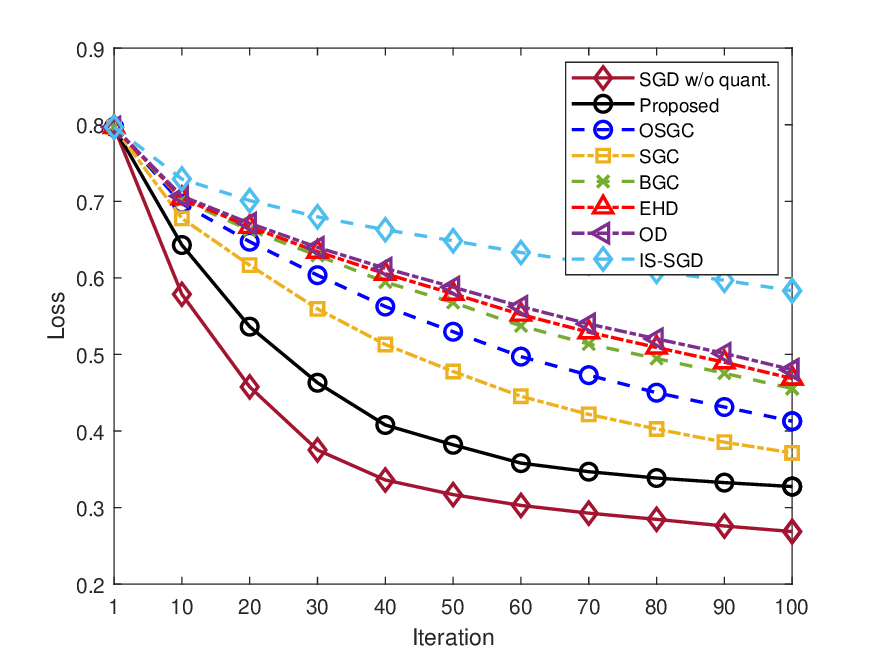}
			\caption{}
			\label{fig:coco1}
		\end{subfigure}
		\begin{subfigure}[b]{0.315\textwidth}
			\includegraphics[width=\linewidth]{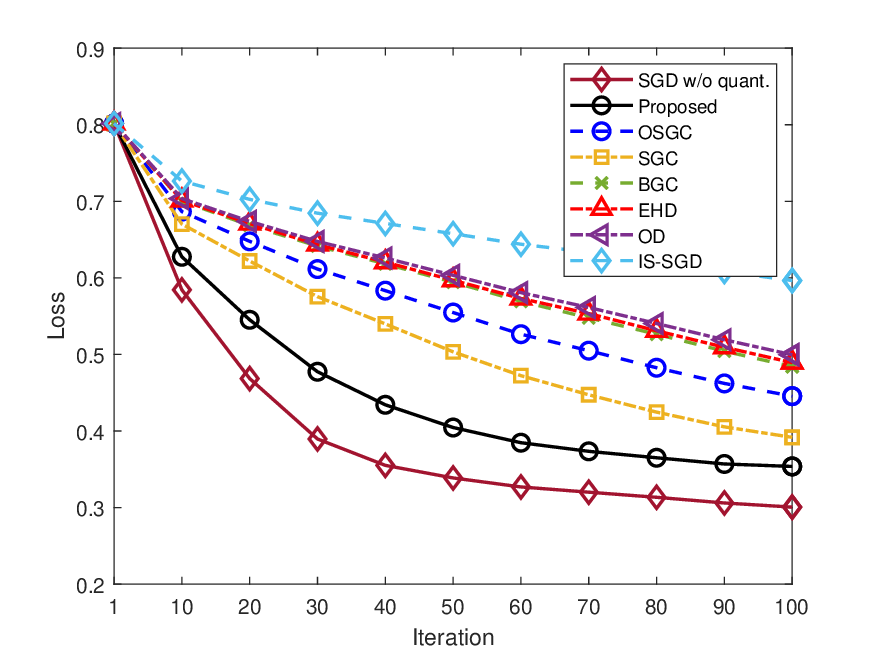}
			\caption{}
			\label{fig:coco3}
		\end{subfigure}
		\begin{subfigure}[b]{0.315\textwidth}
			\includegraphics[width=\linewidth]{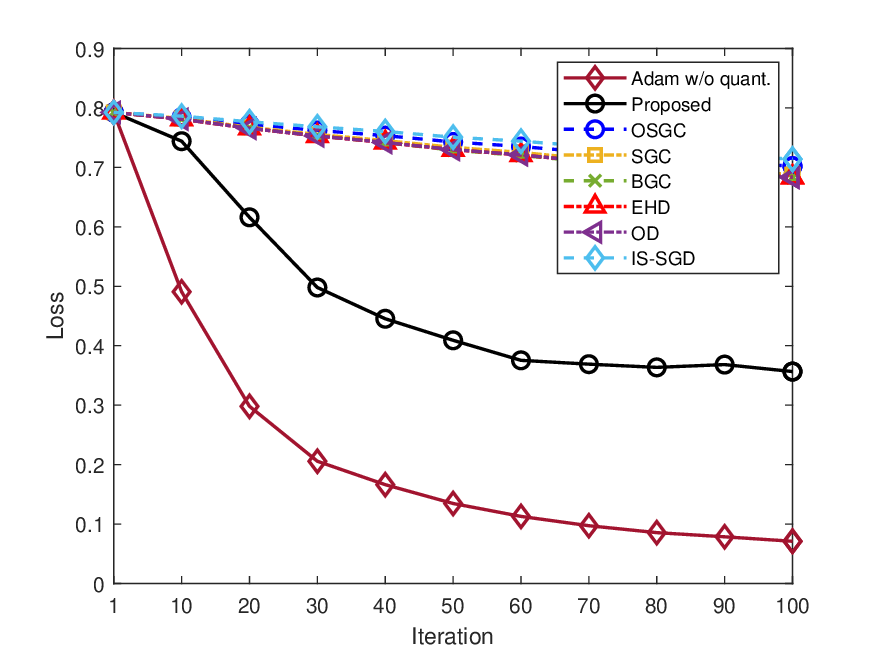}
			\caption{}
			\label{fig:adam1}
		\end{subfigure}
        \\
		\begin{subfigure}[b]{0.315\textwidth}
			\includegraphics[width=\linewidth]{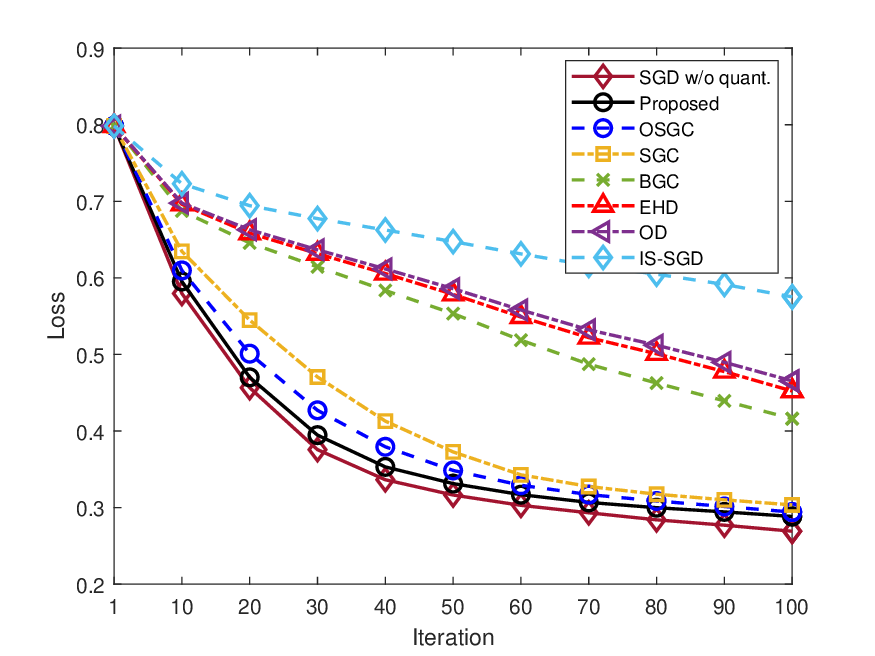}
			\caption{}
			\label{fig:coco2}
		\end{subfigure}
		\begin{subfigure}[b]{0.315\textwidth}
			\includegraphics[width=\linewidth]{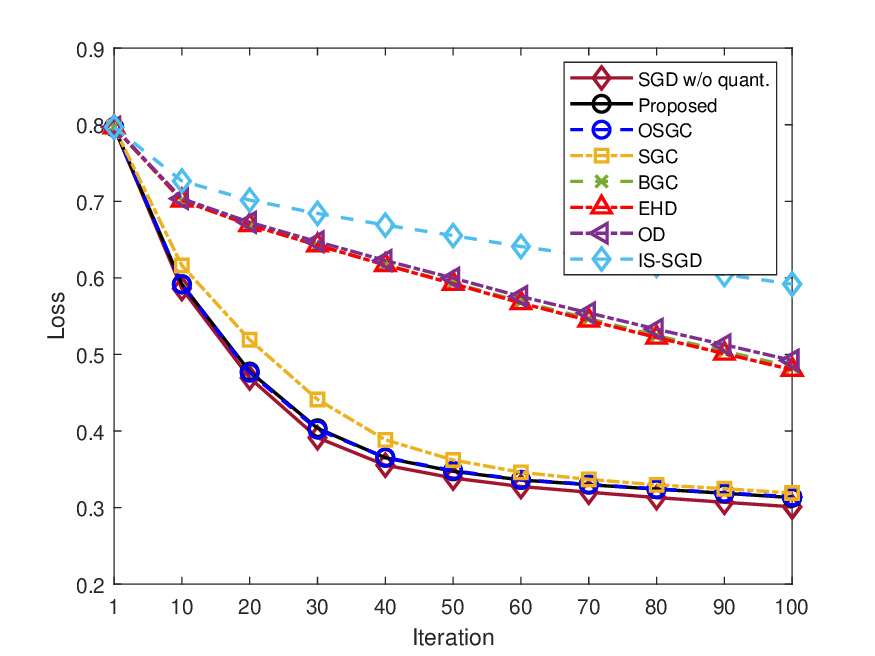}
			\caption{}
			\label{fig:coco4}
		\end{subfigure}
		\begin{subfigure}[b]{0.315\textwidth}
			\includegraphics[width=\linewidth]{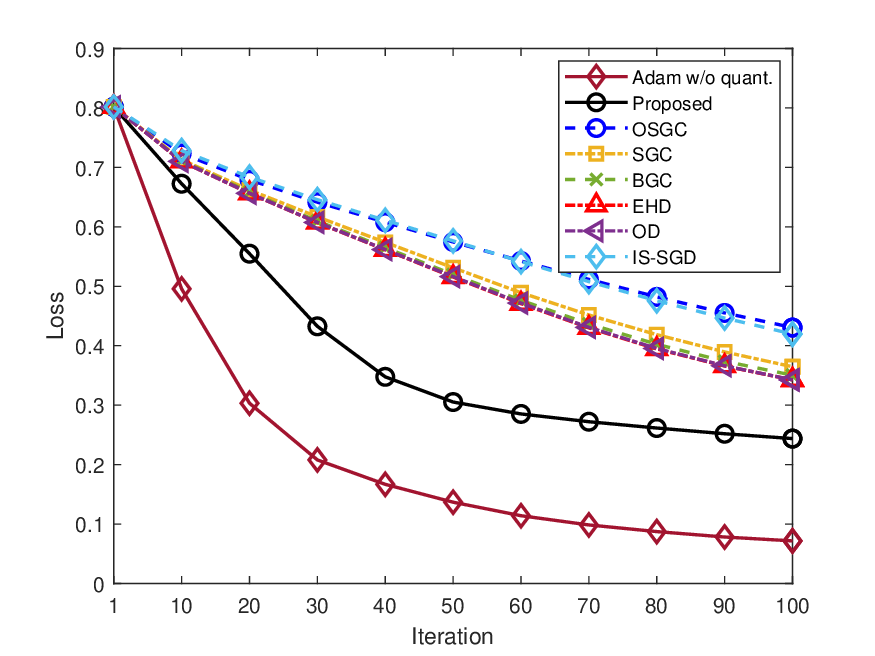}
			\caption{}
			\label{fig:adam2}
		\end{subfigure}
		
		\caption{Convergence graph with respect to the training iteration: bit budget $Z_\text{res}=k$ ((a) SGD $k=10$ (b) SGD $k=50$ (c) Adam $k=10$) and bit budget $Z_\text{res}=5\cdot k$ ((d) SGD $k=10$ (e) SGD $k=50$ (f) Adam $k=10$).}
		\label{fig:conv}
		\vspace{-10pt}
	\end{figure*}
    For our evaluation, we compare our design with benchmarks as follows:
	\begin{itemize}
        \item \textbf{SGD}: Represents an ideal centralized learning scenario where parameters are updated using the true gradient $g^{(t)}$, rendering it immune to straggler delays.
        \item \textbf{IS-SGD} (Ignore-Stragglers SGD): A distributed variant that assigns disjoint data partitions ($d=1$) to workers to eliminate redundancy. However, it lacks mitigation mechanisms and remains vulnerable to straggler effects.
        \item \textbf{BGC \cite{ref:AGC}} (Bernoulli Gradient Coding): Generates encoding coefficients from a Bernoulli distribution, $a_{i,j}\sim \text{Bernoulli}(d/k)$, while fixing $w_i=1, \forall i$.
        \item \textbf{EHD \cite{ref:AGC2}} (ERASUREHEAD): Constructs encoding coefficients based on fractional repetition (FR) codes \cite{ref:GC} and employs uniform decoding coefficients ($w_i=1$).
        \item \textbf{OD \cite{ref:AGC4}} (Optimal Decoding): Utilizes a random graph for encoding ($a_{i,j}\in\{0,1\}$) and dynamically optimizes the decoding coefficients $w_i$ in each iteration to minimize straggler effects.
        \item \textbf{SGC \cite{ref:SGC}} (Stochastic Gradient Coding): Adopts a pair-wise data distribution strategy for unbiased gradient estimation with redundancy defined as in \cite{ref:SGC}. To adapt this method to heterogeneous straggler environments, we modified the encoding coefficients to $a_{i,j} = \frac{1}{d_j(1-p_i)}$ while maintaining $w_i=1, \forall i$. 
        \item  \textbf{OSGC \cite{ref:HAGC}} (Optimally Structured Gradient Coding): OSGC derives coding coefficients via Lagrangian duality to minimize gradient error for heterogeneous stragglers under unbiasedness, without quantization.
	\end{itemize}
    The gradient coding baselines are configured with a computation load of approximately $d\approx 2$, and equal bit allocation is applied as baselines do not consider quantization.

	Fig. \ref{fig:ba} compares the objective value $F(r)$ obtained by our proposed bit allocation algorithm against various benchmarks across varying bit budgets $Z_\text{res}$. We observe that our proposed low-complexity strategy achieves near-optimal performance, consistently matching the global optimum derived from DP, whereas heuristic methods such as greedy and continuous relaxation algorithms frequently become trapped in local optima due to the non-convex nature of the utility function. Notably, this advantage in optimization efficiency persists as the system scales, as evidenced by the consistent performance gap between our method and the baselines in both $k=10$ and $k=50$ scenarios.

     Fig. \ref{fig:conv} demonstrates the convergence behavior of the proposed scheme on the object detection task. In the SGD settings shown in Fig. \ref{fig:conv}(a), \ref{fig:conv}(b), \ref{fig:conv}(d), and \ref{fig:conv}(e), our method consistently surpasses existing gradient coding baselines in terms of both convergence speed and final loss. Notably, this superiority is maintained across different system scales, as evidenced by the consistent performance improvements observed in both $k=10$ and $k=50$ scenarios. It is crucial to highlight that while centralized SGD serves as an ideal baseline for convergence, it suffers from significant time overhead as it relies on a single computing unit. Furthermore, conventional distributed learning typically requires each worker to transmit full-precision gradients (32 bits per coordinate), resulting in an aggregate transmission of $32 \cdot k$ bits per coordinate, which is significantly more costly compared to our total bit budget $Z_\text{tot}$. By jointly considering gradient coding under quantization techniques, our proposed approach effectively addresses these computation and communication bottlenecks. Consequently, it robustly handles heterogeneous stragglers even under tight communication constraints, whereas schemes like OSGC, SGC, EHD, and BGC struggle.  Specifically, OSGC tunes coding coefficients to suppress straggler-induced variance, but without modeling quantization noise these coefficients can inadvertently amplify it, especially under a low bit budget, leading to significant performance degradation.
     Furthermore, our method achieves a convergence trajectory comparable to the ideal centralized scenario, particularly as increasing the bit budget $Z_\text{res}$.

    Moreover, the effectiveness of our design extends to adaptive optimization, as illustrated in the Adam optimizer results in Fig. \ref{fig:conv}(c) and \ref{fig:conv}(f), where the learning rate is set to 0.001. By employing the proposed two-track decoding strategy with $\Lambda=1$, our method achieves significant performance improvements over existing gradient coding techniques, successfully mitigating the estimation bias that typically hinders adaptive optimizers in quantized settings. While two-track decoding empirically improves stability, a full theoretical characterization is deferred to future work.

	\begin{figure*}[t]
		\centering
		\begin{subfigure}[b]{0.315\textwidth} 
			\includegraphics[width=\linewidth]{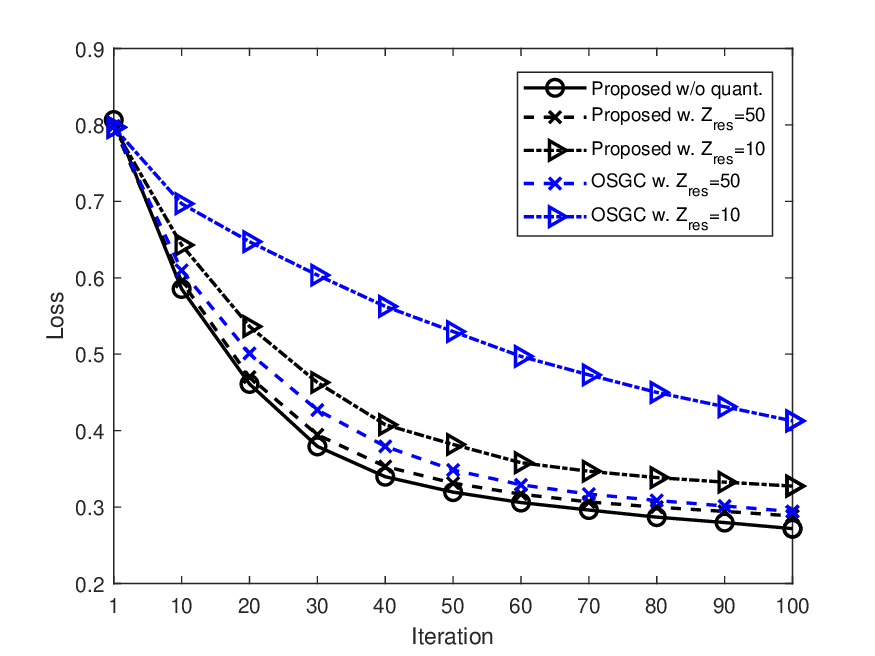}
			\caption{}
			\label{fig:abl}
		\end{subfigure}
		\begin{subfigure}[b]{0.315\textwidth}
			\includegraphics[width=\linewidth]{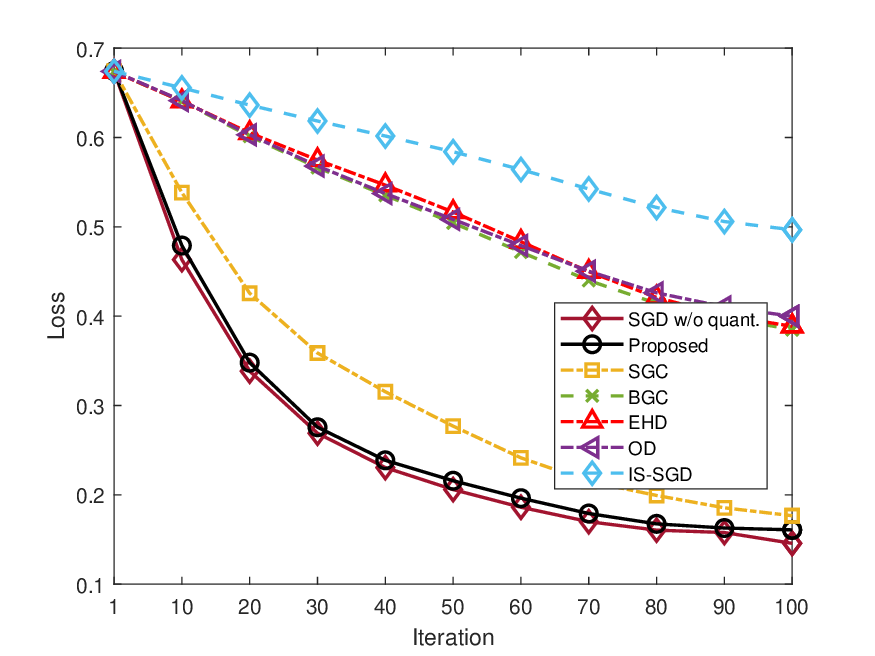}
			\caption{}
			\label{fig:res}
		\end{subfigure}
		\begin{subfigure}[b]{0.315\textwidth}
			\includegraphics[width=\linewidth]{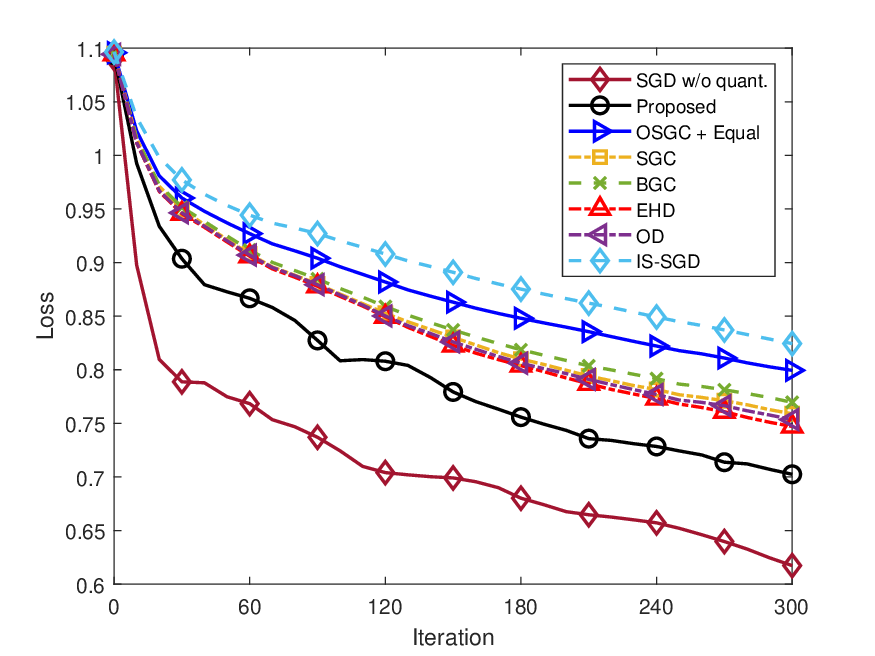}
			\caption{}
			\label{fig:gpt}
		\end{subfigure}
        \\
		\begin{subfigure}[b]{0.315\textwidth}
			\includegraphics[width=\linewidth]{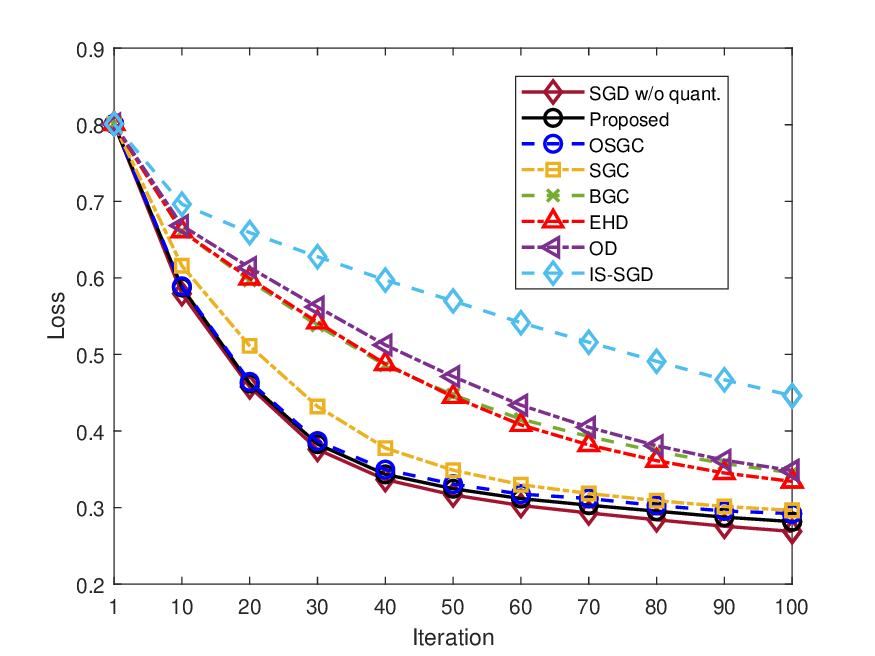}
			\caption{}
			\label{fig:het}
		\end{subfigure}
		\begin{subfigure}[b]{0.315\textwidth}
			\includegraphics[width=\linewidth]{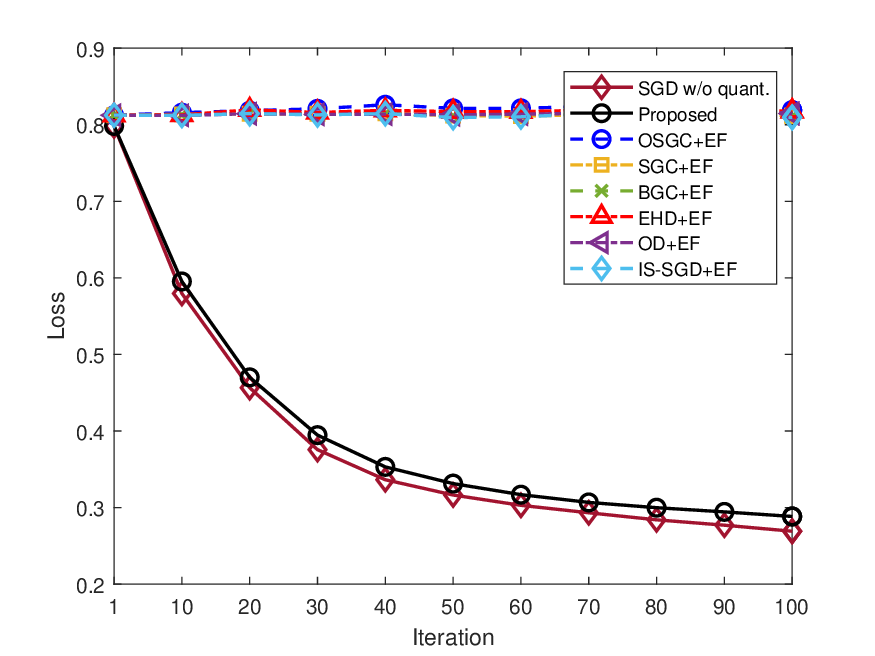}
			\caption{}
			\label{fig:ef}
		\end{subfigure}
		\begin{subfigure}[b]{0.315\textwidth}
			\includegraphics[width=\linewidth]{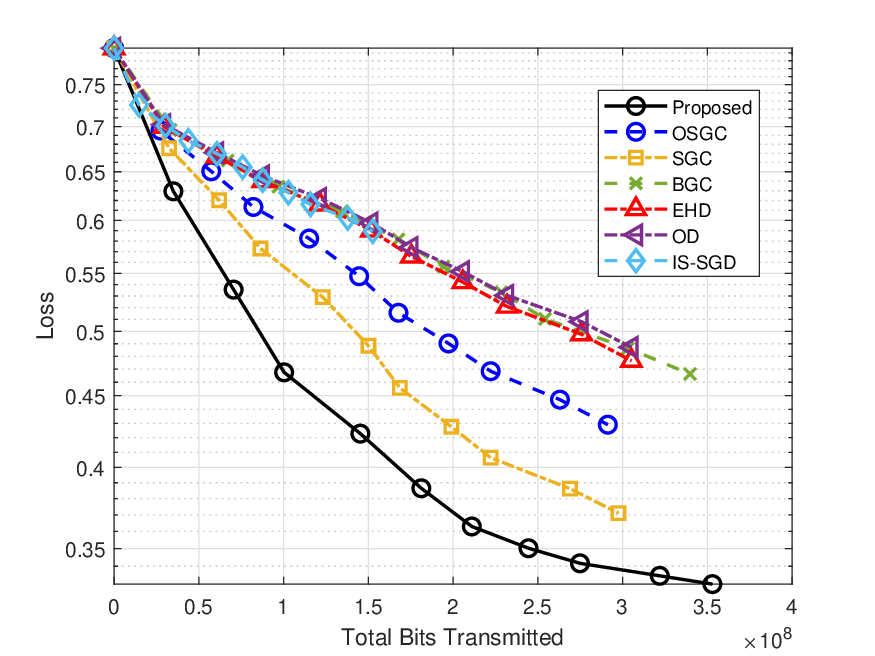}
			\caption{}
			\label{fig:bits}
		\end{subfigure}
		
		\caption{Convergence graph with respect to the training iteration: (a) ablation study (b) ResNet-50 (c) GPT-2 (d) $\mu_{max}=5$ (e) combined with error-feedback method, and (f) with respect to the total bits transmitted.}
		\label{fig:extra}
		\vspace{-10pt}
	\end{figure*}
    Fig. \ref{fig:extra} offers a comprehensive analysis of the proposed scheme's robustness and scalability under various configurations. The ablation study in Fig. \ref{fig:extra}(a) highlights the necessity of the proposed joint design by comparing it with the OSGC \cite{ref:HAGC} baseline with naive quantized extension. Even under a limited bit budget of $Z_{\text{res}}=50$—which provides substantial communication compression compared to the unquantized baseline (${\sim}320$ bits)—the proposed method achieves convergence performance virtually identical to the ideal case. Furthermore, it maintains stability with only a moderate performance drop even under a severely constrained budget of $Z_{\text{res}}=10$. In contrast, the OSGC baseline—which employs equal bit allocation and optimizes solely for straggler mitigation—suffers from significant performance degradation. As evidenced by the inferior loss curves, this instability arises because the coding coefficients, while aggressively tuned to minimize straggler variance, inadvertently amplify the unmodeled quantization noise ($\phi(z_i) > 0$ in Lemma \ref{lem1_quant}), resulting in severe deterioration in convergence behavior. This result critically underscores the importance of our joint optimization strategy, which effectively neutralizes quantization noise through an adaptive design, whereas partial optimization approaches lose robustness, particularly in low-budget regimes. Furthermore, Fig. \ref{fig:extra}(b) demonstrates the scalability of our approach on the ResNet-50 architecture, which contains approximately 25.6 million parameters—roughly 5 times more than the MobileNetV3-Large backbone (${\sim}5.4$M). Even at this increased scale, the experimental insights persist, with our scheme consistently outperforming baselines and maintaining stability.  Finally, Fig. \ref{fig:extra}(c) extends the evaluation to a GPT-2-scale Transformer model, where the proposed method again achieves the best performance among the baselines and remains closest to centralized SGD, confirming that the benefit of the proposed joint design also persists in larger-scale language-model settings.

    Fig. \ref{fig:extra}(d) illustrates the convergence behavior in a highly heterogeneous environment where the upper bound of the straggling parameter is increased to $\mu_{max}=5$. It is worth noting that while a larger $\mu_{max}$ intensifies the degree of heterogeneity across worker nodes, it concurrently reduces the average probability of straggling due to the characteristics of the shifted-exponential model, resulting in generally improved convergence rates for all methods compared to lower $\mu_{max}$ settings. Even under this amplified heterogeneity, the proposed scheme demonstrates superior adaptability and robustness, consistently outperforming state-of-the-art baselines such as SGC and EHD. This confirms that our method effectively leverages the reduced average straggling frequency while successfully mitigating the increased variance in worker capabilities, achieving a convergence trajectory that closely mirrors the ideal centralized SGD. 
        
        We also evaluate error-feedback (EF), a well-established technique for communication-efficient compressed optimization, by applying it to the existing gradient-coding baselines. As shown in Fig.~\ref{fig:extra}(e), the EF-enhanced baselines fail to converge stably, while the proposed method closely tracks SGD. Since EF does not explicitly model straggler events, delayed or missing workers may accumulate stale residuals, which makes residual correction unreliable. By contrast, our method jointly designs the coding structure and bit allocation with both straggling and quantization effects taken into account. 
        Finally, to evaluate end-to-end communication efficiency, we plot the training loss with respect to the total transmitted bits in Fig.~\ref{fig:extra}(f). The proposed method achieves a lower loss than all baselines under the same communication budget. In particular, when the total transmitted bits reach $2.8\times10^8$, the proposed method attains a loss of $0.34$, while SGC and OSGC achieve $0.38$ and $0.44$, respectively. This confirms that the proposed joint design not only improves per-iteration convergence but also reduces the total communication required to reach a given training loss.

	\begin{figure}[!t]	
		\centering	
        
          \begin{subfigure}[t]{0.24\columnwidth}
            \includegraphics[width=\linewidth]{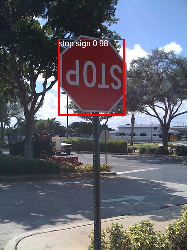}
            \caption{ }
            \label{fig:vis_gd}
          \end{subfigure}
          \begin{subfigure}[t]{0.24\columnwidth}
            \includegraphics[width=\linewidth]{zvis_proposed.eps}
            \caption{ }
            \label{fig:vis_proposed}
          \end{subfigure}
          \begin{subfigure}[t]{0.24\columnwidth}
            \includegraphics[width=\linewidth]{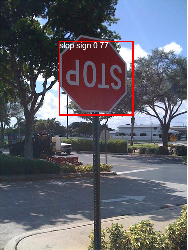}
            \caption{ }
            \label{fig:vis_sgc}
          \end{subfigure}
          \begin{subfigure}[t]{0.24\columnwidth}
            \includegraphics[width=\linewidth]{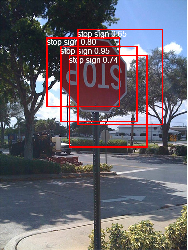}
            \caption{ }
            \label{fig:vis_ehd}
          \end{subfigure}
	\\
          \begin{subfigure}[t]{0.24\columnwidth}
            \includegraphics[width=\linewidth]{zvis_bgc.eps}
            \caption{ }
            \label{fig:vis_bgc}
          \end{subfigure}
          \begin{subfigure}[t]{0.24\columnwidth}
            \includegraphics[width=\linewidth]{zvis_od.eps}
            \caption{ }
            \label{fig:vis_od}
          \end{subfigure}
          \begin{subfigure}[t]{0.24\columnwidth}
            \includegraphics[width=\linewidth]{zvis_issgd.eps}
            \caption{ }
            \label{fig:vis_issgd}
          \end{subfigure}
		\caption{ Detected objects of sampled image: (a) SGD (b) Proposed (c) SGC (d) EHD (e) BGC (f) OD (g) IS-SGD.}
		\label{fig:visual}
	\end{figure}
    Fig. \ref{fig:visual}  presents the object detection results on a sample image from the COCO validation set under a bit budget of $Z_\text{res}=50$. The ideal, centralized learning-based SGD baseline identifies the stop sign with a high confidence score of 0.98. Remarkably, the proposed method achieves an identical score of 0.98, demonstrating performance virtually indistinguishable from the ideal baseline. This contrasts sharply with SGC, which drops to a score of 0.77, and other baselines (EHD, BGC, OD, IS-SGD) that either generate redundant detection boxes or fail to detect the object entirely. These results provide compelling evidence that our optimally structured scheme not only effectively mitigates the impact of stragglers but also significantly reduces the adverse effects of quantization noise, enabling the recovery of high-fidelity gradients even in distributed environments with communication efficiency.

	\section{Conclusion} \label{sec:CONC}	
     We proposed a communication-efficient gradient coding scheme that jointly optimizes straggler mitigation and quantization in heterogeneous distributed systems. By deriving the optimal code structure to achieve a DP-based joint optimum and a near-optimal low-complexity bit allocation, our method minimizes residual error while maintaining unbiasedness. Theoretical and empirical results confirm that our method consistently outperforms benchmarks with low computation and communication loads.  
    Collectively, these results highlight the scheme's practicality for real-world distributed learning scenarios. 

	
	\begin{appendices}

		 \section{Proof of Lemma \ref{lem1_quant}} \label{apdx:lem1} 
        Let $\tilde{g}^{(t)}$ denote the unquantized gradient estimator, i.e., $\tilde{g}^{(t)} = \sum_{i=1}^k \mathbb{I}_i w_i f_i^{(t)}$.  Since the quantization noise is zero-mean and the quantization randomness is independent across workers, the total error variance can be decomposed as:
		\begin{equation}
			\mathbb{E}_{S, Q} [\lVert g^{(t)} - \hat{g}^{(t)} \rVert_2^2] = \underbrace{\mathbb{E}_{S} [\lVert g^{(t)} - \tilde{g}^{(t)} \rVert_2^2]}_{\text{(I) Straggler Error}} + \underbrace{\mathbb{E}_{S, Q} [\lVert \tilde{g}^{(t)} - \hat{g}^{(t)} \rVert_2^2]}_{\text{(II) Quantization Error}}.
		\end{equation}
        
		\textbf{(I) Derivation of Straggler Error:}
		 The detailed algebraic expansion of the straggler error (Term I) follows the same argument as Appendix F.1 of our previous work \cite{ref:HAGC}. 
        Consequently, the straggler error (Term I) is bounded by:
        \begin{equation} \label{eq:straggler_bound}
		\mathbb{E}_{S} [\lVert g^{(t)} - \tilde{g}^{(t)}\rVert_2^2] \leq C \sum_{i=1}^k p_i (1-p_i) \cdot w_i^2 \bigg(\sum_{j=1}^n a_{i,j}\bigg)^2.
		\end{equation}

		\textbf{(II) Derivation of Quantization Error:}
		Next, we analyze the error introduced by stochastic quantization. Since the quantization noise is independent across workers, we have:
		\begin{alignat}{2}
			\mathbb{E}_{S, Q} [\lVert \tilde{g}^{(t)} & - \hat{g}^{(t)} \rVert_2^2] = \mathbb{E}_{S, Q} \bigg[ \bigg\lVert \sum_{i=1}^k \mathbb{I}_i w_i (f_i^{(t)} - Q(f_i^{(t)})) \bigg\rVert_2^2 \bigg]  \nonumber \\
			&= \sum_{i=1}^k (1-p_i) w_i^2 \mathbb{E}_{Q} [\lVert f_i^{(t)} - Q(f_i^{(t)}) \rVert_2^2]  \nonumber \\
			&\leq \sum_{i=1}^k (1-p_i) w_i^2 \phi(z_i) \lVert f_i^{(t)} \rVert_2^2  \nonumber \\
			&\leq C \sum_{i=1}^k (1-p_i) \phi(z_i) w_i^2 \bigg(\sum_{j=1}^n a_{i,j}\bigg)^2, \label{eq:quant_bound}
		\end{alignat}
		where we used the quantization error bound $\mathbb{E}_{Q}[\lVert f_i - Q(f_i) \rVert_2^2] \le \phi(z_i) \lVert f_i \rVert_2^2$ and the gradient boundedness assumption $\lVert f_i^{(t)} \rVert_2^2 \le C (\sum_{j=1}^n a_{i,j})^2$.  
        Specifically, the latter inequality is derived by applying the triangle inequality and Assumption \ref{assumption:const} to the definition of the coded gradient $f_i^{(t)} = \sum_{j=1}^n a_{i,j} g_j^{(t)}$ with $a_{i,j}\ge 0$ as follows:
        \begin{equation}
            \lVert f_i^{(t)} \rVert_2 = \bigg\lVert \sum_{j=1}^n a_{i,j} g_j^{(t)} \bigg\rVert_2 \le \sum_{j=1}^n a_{i,j} \lVert g_j^{(t)} \rVert_2 \le \sqrt{C} \sum_{j=1}^n a_{i,j}.
        \end{equation}
        Squaring both sides yields the stated bound.
		
		\textbf{Total Error Bound:}
		Combining \eqref{eq:straggler_bound} and \eqref{eq:quant_bound}, the total residual error is bounded by:
		\begin{equation}
			\mathbb{E}_{S, Q} [\lVert g^{(t)} - \hat{g}^{(t)}\rVert_2^2] \leq C \sum_{i=1}^k (1-p_i) \{p_i + \phi(z_i)\} \cdot w_i^2 \bigg(\sum_{j=1}^n a_{i,j}\bigg)^2.
		\end{equation}
        
		\section{Proof of theorem \ref{thm:opt_str}} \label{apdx:thm1}
		
		The Lagrangian function of (\textbf{c-SP2}) is 
		\begin{equation}
			\mathcal{L}(\boldsymbol{\alpha}) =\sum_{i=1}^k c_i \bigg( \sum_{j=1}^n  \alpha_i^j \bigg)^2 +\sum_{j=1}^n \zeta_j \bigg(\sum_{i=1}^k \alpha_i^j - 1 \bigg),
		\end{equation}
		where $c_i = \frac{p_i +\phi(z_i)}{1-p_i}$ and $\zeta_j$ is a Lagrangian multiplier. 
		
		By Karush-Kuhn-Tucker (KKT) conditions, we obtain 
		\begin{equation}
			\begin{cases}
				\frac{\partial \mathcal{L}}{\partial \alpha_i^j}=2c_i \bigg( \sum_{j=1}^n \alpha_i^j\bigg) + \zeta_j = 0, \forall i, j,    \\
				\sum_{i=1}^k \alpha_i^j = 1, \forall j.
			\end{cases}
		\end{equation}
		Based on the KKT condition on stationarity, we have 
		\begin{equation}
		c_1 \bigg(\sum_{j=1}^n \alpha_1^j\bigg) = c_2 \bigg(\sum_{j=1}^n \alpha_2^j\bigg) = \cdots = c_n \bigg(\sum_{j=1}^n \alpha_n^j\bigg).
		\end{equation}
        Thus, let $X=c_i \sum_{j=1}^n \alpha_i^j, \forall i$ and using the primal feasibility, i.e., $\sum_{i=1}^k \alpha _i^j =1, \forall j$, we have 
		\begin{equation}
		\sum_{i=1}^k \bigg(X \cdot c_i^{-1}\bigg) = \sum_{i=1}^k \bigg(\sum_{j=1}^n \alpha_i^j \bigg) = n,
		\end{equation}
		and we have, 
		\begin{equation}
		X = \frac{n}{\sum_{i=1}^k c_i^{-1}}.
		\end{equation}
		Consequently, the optimal gradient codes, derived from minimizing the gradient estimation error under the unbiasedness constraint, are computed when the matrix $\boldsymbol{\alpha}$ satisfies the following conditions.
		\begin{equation}
        \sum_{j=1}^n (\alpha_i^j)^* = Y_i, \forall i, \text{ and } \sum_{i=1}^k (\alpha_i^j)^* = 1, \forall j,
        \end{equation}
		where $(\alpha_i^j)^*$ is the optimal $\alpha_i^j$, and  $Y_i=c_i^{-1}\cdot \frac{n}{\sum_{j=1}^k c_j^{-1}}$.
        
        \section{Proof of lemma \ref{lem:ub_err}} \label{apdx:lem_ub_err}
		Based on the Lemma \ref{lem1_quant} and Theorem \ref{thm:opt_str}, we can derive the bounded residual error of the gradient estimator for any optimally structured gradient codes in the following: 
		\begin{alignat}{2}
			\mathbb{E}_{S, Q} [\lVert g^{(t)} - \hat{g}^{(t)}\rVert_2^2] &\leq C \bigg[\sum_{i=1}^k c_i \cdot \bigg( \sum_{j=1}^n (\alpha_i^j)^* \bigg)^2\bigg]  \nonumber \\
			&\leq n^2 C \cdot \frac{1}{\sum_{i=1}^k c_i^{-1}},
		\end{alignat}    
		where $\alpha_i^j = \Tilde{w}_i a_{i,j}$,  $(\alpha_i^j)^*$ represents the optimal $\alpha_i^j$, and $c^{-1}_i = \frac{1-p_i}{p_i + \phi(z_i)}$.
		Furthermore, the squared norm of the gradient estimator for any optimally structured gradient codes is bounded by
		\begin{alignat}{2}
			\mathbb{E}_{S, Q} [\lVert&\hat{g}^{(t)} \rVert_2^2] = 	\mathbb{E}_{S, Q} [\lVert g^{(t)} - \hat{g}^{(t)}\rVert_2^2] + \lVert g^{(t)} \rVert_2^2  \nonumber \\
			&\leq C \bigg[n^2 + \sum_{i=1}^k c_i \cdot \bigg( \sum_{j=1}^n (\alpha_i^j)^* \bigg)^2\bigg]  \nonumber \\
			&=  n^2 C \cdot \bigg( 1+ \frac{1}{\sum_{i=1}^k c_i^{-1}} \bigg).
		\end{alignat}

        \section{Proof of proposition \ref{prop:bitopt}}
        \label{apdx:prop_bitopt}
        Since $\mathbf{z}$ is uniquely determined by adding 2 to $\mathbf{r}$, we present the proof in terms of $\mathbf{z}$. 
        Let $\mathcal{J}(\mathbf{A}, \mathbf{w}, \mathbf{z})$ denote the objective function of the joint optimization problem (\textbf{P2}) as defined in Eq. \eqref{opt:p2}:
        \begin{equation} 
            \mathcal{J}(\mathbf{A}, \mathbf{w}, \mathbf{z}) = \sum_{i=1}^k \frac{p_i +\phi(z_i)}{1-p_i} \tilde{w}_i^2 \bigg( \sum_{j=1}^n a_{i,j} \bigg)^2.
        \end{equation}
        Let $\Omega$ and $\mathcal{Z}$ represent the feasible sets for the coding coefficients $(\mathbf{A}, \mathbf{w})$ and the integer bit allocation $\mathbf{z}$, respectively. The original problem (\textbf{P2}) seeks to find the global minimum. Since the joint feasible set factors as the Cartesian product $\Omega \times \mathcal{Z}$ (i.e., the constraints defining $\Omega$ do not involve $\mathbf{z}$, and the constraints defining $\mathcal{Z}$ do not involve $(\mathbf{A}, \mathbf{w})$), we can apply the principle of partial minimization to rewrite the joint minimization in an exactly equivalent nested form:
        \begin{equation}
            \min_{(\mathbf{A},\mathbf{w},\mathbf{z}) \in \Omega \times \mathcal{Z}} \mathcal{J}(\mathbf{A},\mathbf{w},\mathbf{z}) = \min_{\mathbf{z}\in\mathcal{Z}} \left( \min_{(\mathbf{A},\mathbf{w})\in\Omega} \mathcal{J}(\mathbf{A},\mathbf{w},\mathbf{z}) \right).
        \end{equation}
        Consider the inner minimization for any fixed $\mathbf{z}$.  In this setting, the terms $c_i = \frac{p_i +\phi(z_i)}{1-p_i}$ act as constant scalars; hence the inner minimization optimizes $(\mathbf{A}, \mathbf{w})$ under these weights (i.e., the $\phi(z_i)$-dependent weighting is fully accounted for).
        Consequently, the inner problem is to find the gradient code $(\mathbf{A}, \mathbf{w})$ that minimizes the weighted sum $\mathcal{J}$ subject to the unbiasedness constraint $\sum_{i=1}^k \tilde{w}_i a_{i,j} = 1, \quad \forall j \in [1:n].$ 
        Crucially, this formulation does not imply decoupling the quantization term $\phi(z_i)$ in Eq. \eqref{opt:p2} from the gradient coding variables, nor does it imply minimizing each summand independently. Instead, the inner problem effectively optimizes the gradient code structure specifically adapted to the given bit allocation $\mathbf{z}$. 
        
        Then, this sub-problem is convex with respect to the structural variables $\boldsymbol{\alpha}$ as reformulated in (\textbf{c-SP2}). 
        In Theorem \ref{thm:opt_str} and Lemma \ref{lem:ub_err}, we derived the closed-form optimal value of this inner problem, denoted as $\mathcal{J}^*(\mathbf{z})$:
        \begin{equation}
            \mathcal{J}^*(\mathbf{z}) \triangleq \min_{(\mathbf{A}, \mathbf{w}) \in \Omega} \mathcal{J}(\mathbf{A}, \mathbf{w}, \mathbf{z}) = n^2 C \left( \sum_{i=1}^k h_i(z_i) \right)^{-1},
        \end{equation}
        where $h_i(z_i) = c_i^{-1} = \frac{1-p_i}{p_i + \phi(z_i)}$ represents the utility of worker $i$ dependent on $z_i$.
        
        Since $C$ and $n$ are positive constants and the function $f(x) = x^{-1}$ is strictly monotonically decreasing for $x>0$, minimizing the profile objective $\mathcal{J}^*(\mathbf{z})$ is mathematically equivalent to maximizing the denominator term $\sum_{i=1}^k h_i(z_i)$. The optimization problem (\textbf{b-SP1}) is defined precisely to maximize this sum:
        \begin{equation}
            \mathbf{z}^* = \underset{\mathbf{z} \in \mathcal{Z}}{\arg\max} \sum_{i=1}^k h_i(z_i).
        \end{equation}
        Therefore, solving (\textbf{b-SP1}) identifies the exact $\mathbf{z}^*$ that minimizes the globally optimal error surface derived from the inner problem. This shows that optimizing $\mathbf{z}$ over the profile objective $\mathcal{J}^*(\mathbf{z})$ is exactly equivalent to solving (\textbf{P2}) in a nested form.

        \section{Proof of lemma \ref{lem:dp_opt}}
        \label{apdx:lem_dpopt}
        We prove that the value $V[i, r]$ computed by the recurrence relation \eqref{eq:dp_rec} corresponds to the maximum utility achievable for the subproblem involving the first $i$ workers with a total residual budget $r$. We proceed by induction on $i$.
        
        \textbf{Base Case ($i=1$):} 
        For the first worker, the recurrence is given by $V[1, r] = \max_{0 \le \mathsf{a} \le r} (V[0, r-\mathsf{a}] + h_1(\mathsf{a}))$. Since $V[0, 0]=0$ and $V[0, x]=-\infty$ for $x>0$, the only valid term is when $r-\mathsf{a}=0$, i.e., $\mathsf{a}=r$. Thus, $V[1, r] = h_1(r)$. This is trivially optimal as there is only one worker who must receive the entire allocated budget $r$.
        
        \textbf{Inductive Step:} 
        Assume that for the first $i-1$ workers, $V[i-1, x]$ correctly represents the optimal total utility for any budget $0 \le x \le Z_{res}$. Specifically,
        \begin{equation}
            V[i-1, x] = \max_{\{r_1, \dots, r_{i-1}\}} \sum_{j=1}^{i-1} h_j(r_j) \quad \text{s.t.} \quad \sum_{j=1}^{i-1} r_j = x. 
        \end{equation}
        Now, consider the problem for the first $i$ workers with budget $r$. Let the optimal allocation to the $i$-th worker be $r_i^*$. Then, the remaining budget $r - r_i^*$ must be optimally distributed among the first $i-1$ workers. By the separability of the objective function and the linear constraint, the optimal utility is:
        \begin{equation}
            F^*_i(r) = h_i(r_i^*) + V[i-1, r - r_i^*].    
        \end{equation}
        Since we do not know $r_i^*$ a priori, the recurrence exhaustively searches over all feasible allocations $\mathsf{a} \in [0, r]$ for the $i$-th worker:
        \begin{equation}
        V[i, r] = \max_{0 \le \mathsf{a} \le r} \{ h_i(\mathsf{a}) + V[i-1, r-\mathsf{a}] \}. 
        \end{equation}
        By the inductive hypothesis, $V[i-1, r-\mathsf{a}]$ is the optimal value for the remaining subproblem. Therefore, maximizing the sum over all possible $\mathsf{a}$ guarantees finding the global optimum for $i$ workers.
        
        By induction, $V[k, Z_{res}]$ yields the global maximum of the original problem (\textbf{b-SP1}).

        \section{Proof of proposition \ref{prop:twodec}} \label{apdx:twodec}
        
        We provide the proof for the estimation error guarantee of the proposed second-moment decoder and its stabilization effect on adaptive optimization.
                
        Recall the weighted bias--variance objective for second-moment decoding in \eqref{eq:twodec}:
        \begin{align}
        \mathcal{M}(v) \triangleq & \Lambda \left[\sum_{j=1}^{n}  \Bigg(1-\sum_{i=1}^{k}(1-p_i)v_i a_{i,j}\Bigg)\right]^2  \nonumber \\
    & + \sum_i (1-p_i) \{p_i+\phi(z_i)\}\cdot  v_i^2 \left(\sum_j a_{i,j}\right)^2,
        \end{align}
        where $\Lambda>0$ controls the trade-off between bias-penalty and variance. We define the bias and variance components respectively as:
        \begin{align}
        \mathsf{Bias}(v) &\triangleq \left[\sum_{j=1}^{n} \Bigg(1-\sum_{i=1}^{k}(1-p_i)v_i a_{i,j}\Bigg) \right]^2, \\ 
        \mathsf{Var}(v)
        \triangleq \sum_{i=1}^{k}& (1-p_i) \{p_i+\phi(z_i)\} \cdot v_i^2 \Bigg(\sum_{j=1}^{n} a_{i,j}\Bigg)^2.
        \end{align}
        Thus, the objective can be written as $\mathcal{M}(v)=\Lambda\,\mathsf{Bias}(v)+\mathsf{Var}(v)$.
        
        Let $\mathcal{U}$ be the set of valid unbiased decoders:
        \begin{equation}
        \mathcal{U} \triangleq \Big\{v\in\mathbb{R}^k:\ \sum_{i=1}^{k}(1-p_i)v_i a_{i,j}=1,\ \forall j\in\{ 1,\dots,n\}\Big\}.
        \end{equation}
        For any $w\in\mathcal{U}$, the bias term vanishes (i.e., $\mathsf{Bias}(w)=0$), implying $\mathcal{M}(w)=\mathsf{Var}(w)$.
        
        Let $v^*\in\arg\min_{v\in\mathbb{R}^k}\mathcal{M}(v)$ be the proposed two-track decoder. Since $v^*$ is the global minimizer of $\mathcal{M}(\cdot)$ over the entire space $\mathbb{R}^k$, and the set of unbiased decoders satisfies $\mathcal{U} \subset \mathbb{R}^k$, it follows that $\mathcal{M}(v^*)\le \mathcal{M}(w)$ for any unbiased decoder $w\in\mathcal{U}$. Given that $\mathcal{M}(w)=\mathsf{Var}(w)$ due to the unbiasedness constraint, we establish the following dominance:
        \begin{equation}
        \mathcal{M}(v^*)\le \mathcal{M}(w)=\mathsf{Var}(w).
        \end{equation}
        Furthermore, utilizing the non-negativity of squared bias and variance, we have $\mathsf{Var}(v^*) \le \Lambda\,\mathsf{Bias}(v^*) + \mathsf{Var}(v^*) = \mathcal{M}(v^*) \le \mathsf{Var}(w)$. Similarly, the relation $\Lambda\,\mathsf{Bias}(v^*) \le \mathcal{M}(v^*) \le \mathsf{Var}(w)$ holds. Consequently, $v^*$ guarantees no larger variance and bounded bias compared to any unbiased baseline:
        \begin{equation}
        \mathsf{Var}(v^*)\le \mathsf{Var}(w), \qquad \mathsf{Bias}(v^*)\le \frac{\mathsf{Var}(w)}{\Lambda}.
        \end{equation}

        This guaranteed reduction in the variance of the second-moment estimator has a direct impact on the stability of adaptive optimizers. The Adam optimizer relies on exponential moving averages (EMAs) of the gradient's first and second moments, denoted by $\chi_t$ and $\nu_t$, respectively. The parameter update rule is governed by the ratio of these moments:
        \begin{equation}
        \theta_{t+1} = \theta_t - \gamma \frac{\chi_t}{\sqrt{\nu_t}},
        \end{equation}
        where $\gamma$ is the learning rate. Here, the effective step size is proportional to $\nu_t^{-1/2}$. 
        To analyze the convergence behavior involving the adaptive learning rate, we approximate the expectation of the inverse square root term using a second-order Taylor expansion.
        Consider the function $f(x) = x^{-1/2}$. Its first and second derivatives are $f'(x) = -\frac{1}{2}x^{-3/2}$ and $f''(x) = \frac{3}{4}x^{-5/2}$, respectively.
        Let $x = \nu_t$ and $\bar{\nu} = \mathbb{E}[\nu_t]$. By expanding $f(\nu_t)$ around the mean $\bar{\nu}$, we have:
        \begin{equation}
            \frac{1}{\sqrt{\nu_t}} \approx \frac{1}{\sqrt{\bar{\nu}}} - \frac{1}{2}\bar{\nu}^{-3/2}(\nu_t - \bar{\nu}) + \frac{3}{8}\bar{\nu}^{-5/2}(\nu_t - \bar{\nu})^2.
        \end{equation}
        Taking the expectation on both sides, the linear term vanishes since $\mathbb{E}[\nu_t - \bar{\nu}] = 0$. Recognizing that $\mathbb{E}[(\nu_t - \bar{\nu})^2] = \mathsf{Var}(\nu_t)$, we obtain the following approximation:
        \begin{equation}
            \mathbb{E}\left[\frac{1}{\sqrt{\nu_t}}\right] \approx \frac{1}{\sqrt{\mathbb{E}[\nu_t]}} + \frac{3}{8}(\mathbb{E}[\nu_t])^{-5/2}\mathsf{Var}(\nu_t).
        \end{equation}
        This implies that high-variance fluctuations in the second-moment estimate can lead to step-size inflation because the inverse-square-root mapping is convex, even if the estimator is unbiased.
        By ensuring $\mathsf{Var}(v^*) \le \mathsf{Var}(w)$, the proposed method mitigates this inflation, keeping the step size closer to the optimal scale. Since the variance of the EMA $\nu_t$ is largely governed by the variance of its input sequence (instantaneous squared gradients) through geometric weighting, the reduction in the instantaneous estimation variance via $v^*$ tends to reduce $\mathsf{Var}(\nu_t)$, thereby reinforcing the stabilization effect.

		\section{Proof of theorem \ref{thm:sc}} \label{apdx:thm2}
		Our proof builds upon the result from \cite{ref:SC}, which demonstrates that any algorithm utilizing an unbiased estimator of the true gradient achieves a convergence rate of $O(1/T)$:
		\begin{lemma} (Lemma 1 in \cite{ref:SC}) \label{lem3}
			Suppose the loss function is $\lambda$-strongly convex and the gradient estimator is unbiased. Furthermore, assume $\mathbb{E}_{S, Q} [\lVert\hat{g}^{(t)} \rVert_2^2] \leq G$. Then, by setting $\gamma_t=1/(\lambda t)$, the following holds for any $T$ that 
			\begin{equation}
			\mathbb{E} [\lVert \boldsymbol{\beta}_T - \boldsymbol{\beta}^* \rVert^2_2] \leq \frac{4G}{\lambda T}.
			\end{equation}
		\end{lemma}
		
		Building on the result from Lemma \ref{lem:ub_err}, we can conclude that 
		\begin{equation}
		\mathbb{E}_{S, Q} [\lVert\hat{g}^{(t)} \rVert_2^2] \leq n^2 C  \bigg( 1+ \frac{1}{\sum_{i=1}^k c_i^{-1}} \bigg).
		\end{equation} 
		Thus, by replacing $G$ with the right-hand side of the above inequality, Lemma \ref{lem3} yields 
		\begin{equation}
		\mathbb{E} [\lVert \boldsymbol{\beta}_T - \boldsymbol{\beta}^* \rVert^2_2]\leq \frac{4 n^2 C}{\lambda^2 T}  \bigg(1+\frac{1}{\sum_{i=1}^k c^{-1}_i} \bigg),
		\end{equation}
		where $c^{-1}_i=\frac{1-p_i}{p_i+\phi(z_i)}$.

		\section{Proof of theorem \ref{thm:s}} \label{apdx:thm3}
		From the property of $\mu$-smoothness, 
		\begin{alignat}{2}
			&L(\boldsymbol{\beta}_{t+1}) = L(\boldsymbol{\beta}_{t}-\gamma_t \cdot \hat{g}^{(t)}) \nonumber  \\
			&\leq L(\boldsymbol{\beta}_{t}) - \langle g^{(t)}, \gamma_t \cdot \hat{g}^{(t)} \rangle+\frac{\mu \gamma_t^2 }{2} \lVert\hat{g}^{(t)} \rVert^2_2.
		\end{alignat}  
		By taking the expectation $\mathbb{E}_{S, Q}[\cdot]$ conditioned on the previous iteration on both hand sides, we have
		\begin{alignat}{2}
			&\mathbb{E}_{S, Q} [L(\boldsymbol{\beta}_{t+1})] \nonumber \\
            &\leq L(\boldsymbol{\beta}_{t}) - \langle  g^{(t)}, \gamma_t \cdot \mathbb{E}_{S, Q} [\hat{g}^{(t)}] \rangle+\frac{\mu  \gamma_t^2 }{2}\mathbb{E}_{S, Q} [ \lVert  \hat{g}^{(t)} \rVert^2_2] \nonumber 
            \\
			& \overset{\text{(a)}}\leq L(\boldsymbol{\beta}_{t}) - \gamma_t \cdot \lVert g^{(t)} \rVert^2_2 + \frac{\mu  \gamma_t^2 n^2 C}{2}   \bigg(1+\frac{1}{\sum_{i=1}^k c^{-1}_i} \bigg),
		\end{alignat}
		where (a) comes from the unbiasedness of gradient estimator and Lemma \ref{lem:ub_err}. Taking full expectation $\mathbb{E}[\mathbb{E}_{S, Q}[\cdot]]$ on both sides and rearranging, we obtain
		\begin{alignat}{2}
			 \gamma_t \cdot \mathbb{E} [\lVert g^{(t)} \rVert^2_2] \leq &\mathbb{E} [L(\boldsymbol{\beta}_{t})] - \mathbb{E} [L(\boldsymbol{\beta}_{t+1}) ]  \nonumber  \\
            &+ \frac{\mu  \gamma_t^2 n^2 C}{2}  \bigg(1+\frac{1}{\sum_{i=1}^k c^{-1}_i} \bigg).
		\end{alignat}
		Based on this inequality, we have
		\begin{alignat}{2}
			& \sum_{t=0}^T \gamma_t \cdot \mathbb{E} [\lVert g^{(t)} \rVert^2_2]  \nonumber \\
            &\leq L(\boldsymbol{\beta}_{0}) - \mathbb{E} [L(\boldsymbol{\beta}_{T+1}) ] + \frac{\mu   n^2 C}{2}  \bigg(1+\frac{1}{\sum_{i=1}^k c^{-1}_i} \bigg)\sum_{t=0}^T \gamma_t^2 \nonumber \\
			&\quad \overset{\text{(b)}}\leq L(\boldsymbol{\beta}_{0}) - L(\boldsymbol{\beta}^*) + \frac{\mu   n^2 C}{2}  \bigg(1+\frac{1}{\sum_{i=1}^k c^{-1}_i} \bigg)\sum_{t=0}^T \gamma_t^2,
		\end{alignat}
		where (b) is due to $\mathbb{E} [L(\boldsymbol{\beta}_{T+1})] \geq L(\boldsymbol{\beta}^*)$.
		
		If the learning rate is fixed, i.e., $\gamma_t=\gamma = 1/(T+1)^{1/2}$, we have
		\begin{align}
			\frac{1}{T+1} \sum_{t=0}^T & \mathbb{E} [\lVert g^{(t)} \rVert^2_2] \leq  \frac{L(\boldsymbol{\beta}_{0}) - L(\boldsymbol{\beta}^*)}{(T+1)^{1/2}}  \nonumber \\
            &+\frac{1}{(T+1)^{1/2}}\frac{\mu   n^2 C}{2}  \bigg(1+\frac{1}{\sum_{i=1}^k c^{-1}_i} \bigg),
		\end{align}
		where the following limit holds:
		\begin{equation}
			\lim_{T\rightarrow\infty} \frac{1}{T+1} \sum_{t=0}^T \mathbb{E} [\lVert g^{(t)} \rVert^2_2] =  0.
		\end{equation}
		
		Moreover, if the learning rate is decaying, i.e., $\gamma_t={1}/{(t+1)^{1/2}}$, we have
		\begin{equation}
		\sum_{t=0}^T \frac{1}{(t+1)^{1/2}} \cdot \mathbb{E} [\lVert g^{(t)} \rVert^2_2] \geq  \frac{1}{(T+1)^{1/2}} \sum_{t=0}^T  \mathbb{E} [\lVert g^{(t)} \rVert^2_2].
		\end{equation}
		Thus, using this relations, 
        \begin{align}
        &\frac{1}{T+1}\sum_{t=0}^T \mathbb{E}\!\left[\lVert g^{(t)} \rVert_2^2\right] \nonumber \\
        &\leq \frac{L(\boldsymbol{\beta}_{0})-L(\boldsymbol{\beta}^*)}{(T+1)^{1/2}} + \frac{\mu n^2 C /2}{(T+1)^{1/2}} \left(1+\frac{1}{\sum_{i=1}^k c_i^{-1}}\right) \sum_{t=0}^T \frac{1}{t+1} \nonumber \\
        &\overset{\text{(c)}}{\leq} \frac{L(\boldsymbol{\beta}_{0})-L(\boldsymbol{\beta}^*)}{(T+1)^{1/2}} \nonumber \\
        &\quad+ \frac{\mu n^2 C\left(1+\log(T+1)^{1/2}\right)}{(T+1)^{1/2}} \left(1+\frac{1}{\sum_{i=1}^k c_i^{-1}}\right).
        \end{align}
		where (c) is due to the fact $\sum_{t=0}^T \frac{1}{t+1}\leq 2+\log(T+1)$. Since $\lim_{x\rightarrow\infty} \frac{\log x}{x}=0$, the following limit holds:
		\begin{equation}
			\lim_{T\rightarrow\infty} \frac{1}{T+1} \sum_{t=0}^T \mathbb{E} [\lVert g^{(t)} \rVert^2_2]  = 0.
		\end{equation}
        
	\end{appendices}


\end{document}